\documentclass[superscriptaddress,aps,pra,showpacs,twoside,twocolumn,10pt]{revtex4-1}
\usepackage[colorlinks=true, citecolor=red, urlcolor=blue ]{hyperref}
\usepackage{epsfig,newlfont,amssymb,amsfonts,amsmath,bm,subfigure,palatino,mathtools,amsthm,braket,times,soul,enumitem,color}
\usepackage[normalem]{ulem}

\usepackage{graphicx,cancel}
\usepackage{bbold}
\usepackage{multirow}
\usepackage[table,xcdraw]{xcolor}
\usepackage{array}
\newcolumntype{C}[1]{>{\centering\arraybackslash}m{#1}}
\newtheorem{definition}{Definition.}
\newtheorem{proposition}{Proposition.}
\newtheorem{theorem}{Theorem.}
\begin{document}

\title{Design and Experimental Performance of Local Entanglement Witness Operators}

\author{David Amaro}
\affiliation{Department of Physics, Swansea University, Singleton Park, Swansea SA2 8PP, United Kingdom}

\author{Markus M\"{u}ller}
\affiliation{Department of Physics, Swansea University, Singleton Park, Swansea SA2 8PP, United Kingdom}
\affiliation{Institute for Quantum Information, RWTH Aachen University, D-52056 Aachen, Germany}
\affiliation{Peter Gr\"{u}nberg Institute, Theoretical Nanoelectronics,
Forschungszentrum J\"{u}lich, D-52425 J\"{u}lich, Germany}

\vspace{-3.5cm}

\date{\today}

\begin{abstract}
	Entanglement is a central concept in quantum information and a key resource for many quantum protocols. In this work we propose and analyze a class of entanglement witnesses that detect the presence of entanglement in subsystems of experimental multi-qubit stabilizer states. The witnesses we propose can be decomposed into sums of Pauli operators and can be efficiently evaluated by either two measurement settings only or at most a number of measurements that only depends on the size of the subsystem of interest. We provide two constructive methods to design the local witness operators, the first one based on the local unitary equivalence between graph and stabilizer states, and the second one based on sufficient and necessary conditions that the respective set of constituent Pauli operators needs to fulfill. We theoretically establish the noise tolerance of the proposed witnesses and benchmark their practical performance by analyzing the local entanglement structure of an experimental seven-qubit quantum error correction code.
\end{abstract}

\maketitle

\begin{figure*}[t]
	\centering
	\includegraphics[width=2\columnwidth]{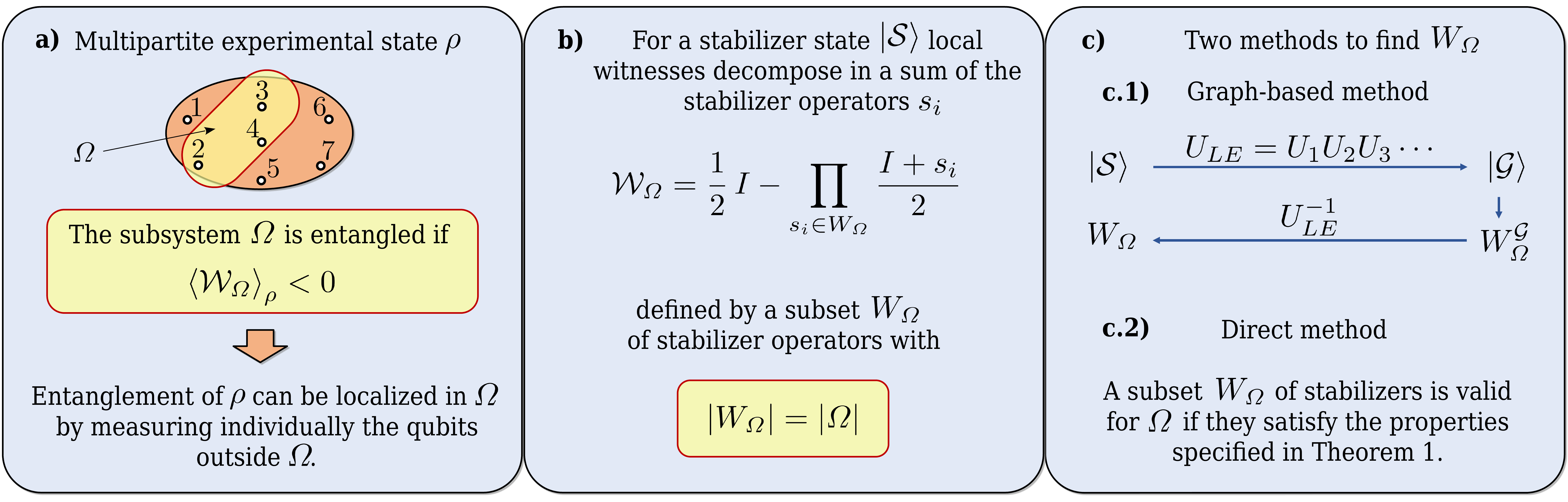}
	\caption{\textbf{Schematic summary of features and construction techniques of local entanglement witnesses.} \textbf{(a)} Given an experimental multipartite state $ \rho $, if the measurement of a local witness $ \mathcal{W}_{\varOmega} $ yields a negative expectation value, a non-zero amount of entanglement can be localized in $ \varOmega $ by local measurements on the qubits outside $ \varOmega $. \textbf{(b)} A local witness constructed for a stabilizer state $ \ket{\mathcal{S}} $ can be decomposed into a sum of stabilizer elements $ s_{i} $ defined by a subset $ W_{\varOmega}\subset\mathcal{S} $ of the stabilizer group $\mathcal{S} $. Importantly, the number of stabilizers in $ W_{\varOmega} $ equals the number of qubits in $ \varOmega $. \textbf{(c)} We propose two constructive methods to find valid sets $ W_{\varOmega} $. \textbf{(c.1)} The \textit{graph-based method} consists in first finding a local unitary $ U_{LE} $ (a product of single-qubit unitaries) that maps the stabilizer state of interest onto a locally equivalent graph state, $ U_{LE}\ket{\mathcal{S}}=\ket{\mathcal{G}} $. Then, the subset $ W_{\varOmega}^{\mathcal{G}}=\left\lbrace g_{\mu}:\;\mu\in\varOmega\right\rbrace  $ of graph state generators that constructs a local witness for $ \ket{\mathcal{G}} $ is identified, and the inverse transformation yields the generator $ W_{\varOmega}=U_{LE}^{-1}W_{\varOmega}^{\mathcal{G}}U_{LE} $. \textbf{(c.2)} The \textit{direct method} consists in checking the four conditions summarized in Theorem \ref{p}, which the stabilizers $ s_{i}\in W_{\varOmega} $ must satisfy to construct a valid local witness for $ \varOmega $.}
	\label{Local_entanglement}
\end{figure*}

\section{Introduction}

Entanglement is one of the fundamental aspects of quantum physics. It is used as a resource, for instance in quantum communication protocols \cite{Gisin2002,Gianfranco2015} or in measurement-based quantum computing \cite{Nielsen2006,Briegel2009}. From a fundamental viewpoint, the presence of entanglement allows one to rule out certain local realistic descriptions of Nature \cite{Buhrman2010,Brunner2014}. Recently, entanglement has also moved into the focus of other research areas beyond the field of quantum information science: examples include studies about the role of entanglement in quantum phase transitions \cite{Vidal2003,Amico2008,Jurcevic2017,Zayed2017}, the presence of long-ranged quantum correlations as a signature of topologically ordered states in condensed matter systems \cite{Kitaev2006, Levin2006,Chen2010, Jahromi2017,DeChiara2018}, or in the AdS/CFT correspondence, where the entanglement entropy in a conformal field theory contains information about the spacetime geometry of the anti-de Sitter space \cite{Ryu2006,Almheiri2015,Pastawski2015,Hubeny2015}.

Experimentally, a variety of physical systems including trapped ions \cite{Brown2016}, photons \cite{Stefanie2015}, cold atoms \cite{Bloch2008}, or superconducting qubits \cite{Clarke2008}, are used to create complex multi-qubit quantum states where entanglement can be studied from a fundamental point of view \cite{Li2011, Larsson2014}, or used as a resource for quantum communication \cite{Gisin2002,Gianfranco2015}, computation \cite{Ladd2010} and simulation \cite{Georgescu2014}. In these systems the study of entanglement through tomographic techniques like compressed sensing \cite{Gross2010, Flammia2012, Steffens2017} is feasible for small systems \cite{Riofrio2017} but becomes impractical as the number of qubits in the systems increases. Two common approaches to overcome this difficulty are MPS tomography \cite{Cramer2010, Ohliger2013, Zhao2017, Lanyon2017}, which efficiently reconstructs the state of systems close to matrix product states, and entanglement witnesses \cite{Terhal2002, Horodecki2009, Guehne2009} which are observables that detect the presence of entanglement with a reduced number of measurements.

Entanglement witnesses have been developed for diverse scenarios, including multi-qubit states \cite{Wieczorek2009, Li2011, Pu2018}, continuous variable systems \cite{Horodecki2000, Hyllus2006, Zhang2013}, thermal states \cite{Toth2005_71, Souza2009, Ribeiro2012, Chakraborty2013}, high-dimensional states \cite{Krenn2014, Huang2016, Ding2016, Howland2016}, or as a way of not only detecting, but also quantifying the amount of entanglement \cite{Brandao2005, Eisert2007, Guehne2008, Cianciaruso2016}. In the context of quantum communications, measurement-device-independent witnesses \cite{Branciard2013, Xu2014, Nawareg2015, Verbanis2016} can be used to prevent eavesdropping by certifying entanglement beyond measurement imperfections.

With respect to the construction of optimal witness operators, work has mainly focused on the decomposition of witnesses into Pauli operators \cite{Bourennane2004, Toth2005_72}, or on reducing the number of required measurement settings \cite{Toth2005_94, Shahandeh2017}. Most efforts are devoted to the detection of genuine entanglement, but there are also witnesses that detect the entanglement depth \cite{Yeong2015, Zhao2016}, the entanglement with respect to partitions \cite{Sperling2013}, or witnesses that provide information about the Schmidt number \cite{Sperling2011, Huber2013, Shahandeh2014}.

In this work we focus on \textit{local witness operators} \cite{Alba2010,Nicolai2018,Amaro2018}, i.e.~witnesses that detect the entanglement among qubits inside subsets $ \varOmega $ of larger multi-qubit systems (see Fig.~\ref{Local_entanglement}(a)). As we will show, the entanglement detected by a local witness is intimately related to \textit{localizable entanglement} \cite{Verstraete2004,Popp2005,Sadhukhan2015}, which is the maximum entanglement that can be localized in a region (subset of qubits) $ \varOmega $ by means of single-qubit projective measurements on the qubits outside $ \varOmega $. In fact, we show that if a local witness detects entanglement in a subsystem $ \varOmega $, the localizable entanglement in $ \varOmega $ is nonzero.

The use of multiple local witnesses for multiple regions $ \varOmega $ reveals the existing entanglement structure of experimental states. With this information it is possible to answer e.g.~whether qubits in a given region or pairs of qubits are entangled. It also allows one to study the entanglement of the subsystem of interest coupled to an environment represented by the rest of qubits. In the preparation of complex many-qubit quantum states, such information may be useful to detect in which spatial regions and within which subsets of particles errors have occurred.

The local witnesses we propose are constructed for stabilizer states and represent a generalization of the local witnesses proposed in \cite{Alba2010} for graph states. Stabilizer states play a role in many areas of quantum information, e.g.~in quantum error correction, where fragile quantum information of logical qubits is distributed over many physical qubits and collectively encoded in entangled stabilizer quantum error correcting codes \cite{Nielsen2010, Devitt2013}. We show that, like genuine entanglement witnesses \cite{Bourennane2004}, local witnesses for stabilizer states can be decomposed into sums of Pauli operators, which belong to the stabilizer group that define the state of interest (Fig.~\ref{Local_entanglement}(b)).

In order to find valid decompositions of local witnesses into sets $ W_{\varOmega} $ of stabilizer operators (see Fig.~\ref{Local_entanglement}(b)) we propose two methods, which will be called the \textit{\textbf{graph-based method}} and the \textit{\textbf{direct method}} in the remainder of the paper. The graph-based method (see Figure \ref{Local_entanglement}(c.1)) consists of finding multiple graph states and the local unitary operations (product of single-qubit unitary operators) connecting them, then constructing multiple local witness for each graph state, and finally transforming them into local witnesses for the stabilizer state via the inverse unitary operations. It was shown in \cite{Van2004} that for every stabilizer state it is always possible to find such local unitary operations and locally equivalent graph states. The direct method resumed in Fig.~\ref{Local_entanglement}(c.2) establishes sufficient and necessary criteria that the stabilizers in $ W_{\varOmega} $ must satisfy to guarantee that it can be transformed into a local witness of some graph state with only local unitary operations.

Apart from providing information about the entanglement in subsets of qubits, the number of measurement settings required to evaluate the local witnesses does not depend on the total number of qubits in the state. Therefore, they can be applied efficiently to states with an in principle arbitrarily large number of qubits. In order to reduce even further the number of measurements required we use the techniques in \cite{Toth2005_94} to propose two types (Fig.~\ref{Local_entanglement}(c)) of modified local witnesses that require even less measurement settings. These techniques are based on reducing the number of stabilizers that appear in the decomposition of the witness while keeping the ones that can be measured with the same measurement setting. For instance, with only two measurement settings it is possible to evaluate all the modified local witness of one type for all the subsystems of the experimental state. The downside of reducing the number of measurements is that the tolerance to noise of the witness decreases. To analyze this quantitatively, we benchmark the performance of the witnesses for noisy quantum states, described a white noise model as previously used e.g.~in \cite{Guehne2009}. We also study the tendency of some witnesses to have a more negative expectation value than others, which means that they are finer entanglement detectors \cite{Lewenstein2000}.

To test the performance of the proposed local witnesses in practice, we use both methods to construct multiple local witnesses for multiple subsystems (Fig.~\ref{Local_entanglement}(f)) of a seven-qubit error correcting code. The analyzed state is a quantum error correction code corresponds to a minimal instance of topological color code, and encodes one protected logical qubit in entangled states of seven physical qubits \cite{Bombin2006}. Additionally, we use experimental data from a recent experimental realization \cite{Nigg2014} of this state to evaluate all local entanglement witnesses put forward, and thereby shed light on the local entanglement structure of the experimental state.

In Section \ref{S2} we revise the concept of genuine entanglement witness operators and particularize them to stabilizer states. In Section \ref{S3}, we introduce the local witnesses for stabilizer states, propose modified local witnesses that require a reduced number of measurement settings, and compare their respective noise tolerance. Section \ref{S4} describes the two methods to construct these operators. In Section \ref{S5}, we present the results of the construction of local witnesses for the seven-qubit color code and then apply them to the experimental realization of this state to map out its entanglement structure. Section \ref{S6} presents our conclusions and an outlook. 

\section{Genuine entanglement witness operators}\label{S2}
Entanglement witnesses are observables that provide a sufficient (thought not necessary) condition for the presence of entanglement. They detect entanglement in noisy experimental quantum states, as long as the difference with the ideal state is sufficiently small. In this section we revise the concept of genuine entanglement witness operators and how to construct them for stabilizer states as explained in \cite{Bourennane2004,Guehne2009}. Evaluation of these witnesses requires a number of measurements that grows in general exponentially with the number of qubits in the state. This motivates us to apply the methods from \cite{Toth2005_94,Toth2005_72} to reduce the required number of measurement settings. 

\subsection{Witnesses as entanglement detectors}\label{genuine_detectors}
An entanglement witness that detects the genuine $ N $-qubit entanglement is an observable which is guaranteed to have a non-negative expectation value if applied to any separable state, and a negative expectation value for at least one genuine entangled state \cite{Terhal2002}. Therefore, a negative expectation value unambiguously signals the presence of $ N $-partite genuine entanglement. A witness $ \mathcal{W} $ prepared for the ideal expected state $ \ket{\psi} $ is an operator
\begin{equation}\label{a}
\mathcal{W}=\alpha_{\ket{\psi}}I-\ket{\psi}\bra{\psi},
\end{equation}
where $ \ket{\psi} $ is a non-separable pure state and $\alpha_{\ket{\psi}}$ is the maximal Schmidt coefficient of all bi-partitions of $ \ket{\psi} $ \cite{Bourennane2004}.

Note that the witness expectation value is directly related to the quantum state fidelity $ F=\bra{\psi}\rho\ket{\psi} $ by $ \braket{\mathcal{W}}_{\rho}=\alpha_{\ket{\psi}}-F $, and therefore a sufficiently high quantum state fidelity suffices to signal the presence of entanglement. For many states the state fidelity can be either estimated efficiently \cite{Flammia2011} or be directly determined \cite{Bourennane2004} via a few measurement settings. For stabilizer states $ \ket{\mathcal{S}} $, which we focus on here, the projector $ \ket{\mathcal{S}}\bra{\mathcal{S}} $ can be decomposed into a sum of the stabilizers that define the state.

\subsection{Stabilizer states}
A \textit{stabilizer state} $ \ket{\mathcal{S}} $ \cite{Gottesman1997, Nielsen2010} is an $ N $-qubit state completely defined by $ N $ independent generators $ s_{i} $ (the generator set), which mutually commute and act trivially on the state, $ \ket{\mathcal{S}}=s_{i}\ket{\mathcal{S}} $. The Pauli operators $ s_{i} $ are tensor products of $ N $ single-qubit Pauli operators $ \left\lbrace I,X,Y,Z\right\rbrace $ multiplied by factors $ \pm1,\pm i $. We denote the usual single-qubit Pauli operators and the identity as $ X $, $ Y $, $ Z $ and $ I $.
\begin{figure}[t]
	\centering
	\includegraphics[width=1\columnwidth]{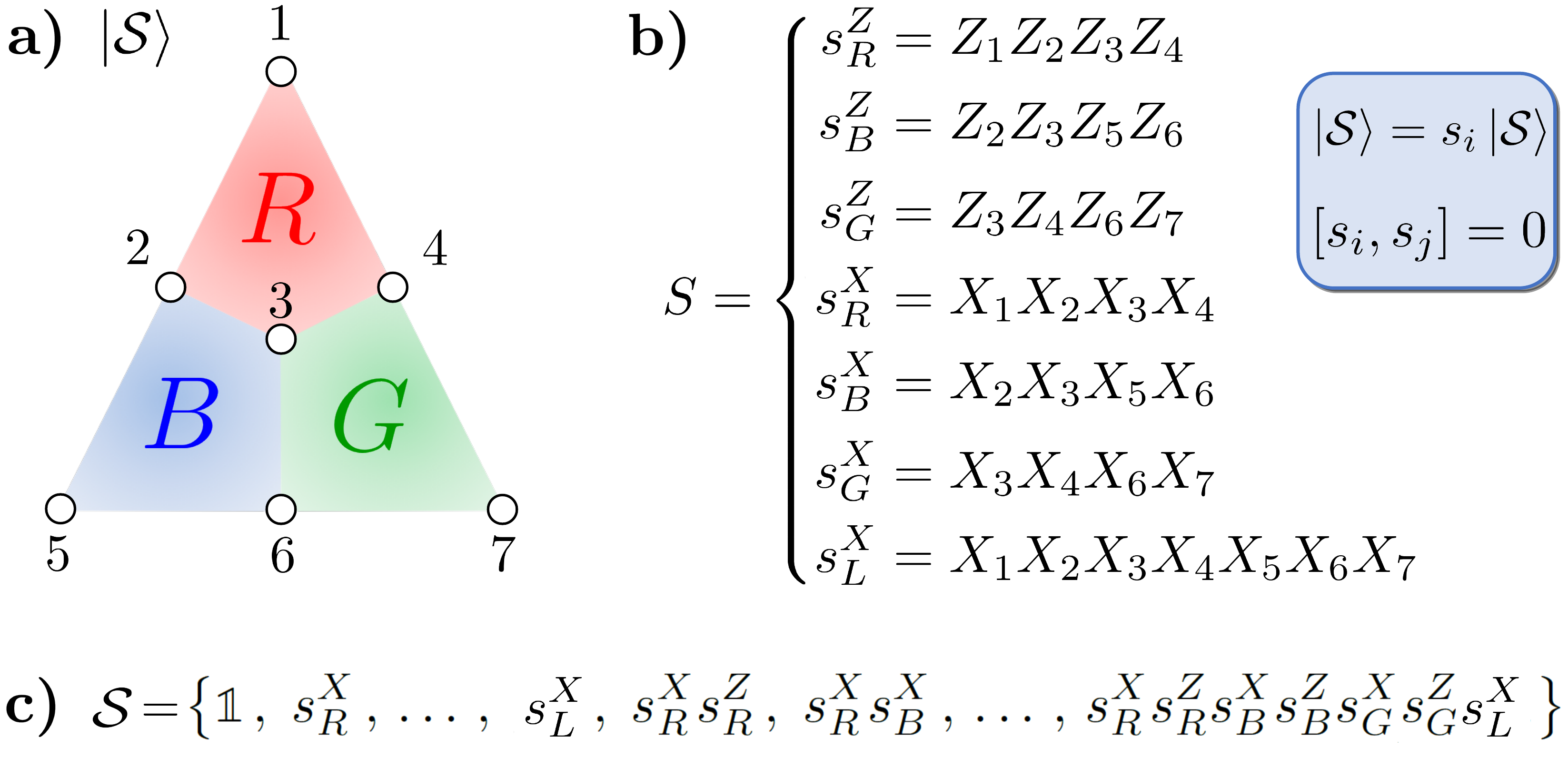}
	\caption{\textbf{Example of a stabilizer state.} \textbf{(a)} Seven-qubit color code lattice containing 7 qubits in the vertices of 3 plaquettes colored in red $ R $, blue $ B $ and green $ G $ \cite{Bombin2006}. \textbf{(b)} \textit{Generator set} $ S $ containing all the \textit{generators} $ s_{i} $ (and their products) that completely define the stabilizer state $ \ket{\mathcal{S}} $, which in this example is the logical state $ \ket{+_{L}} $. \textbf{(c)} \textit{Stabilizer} $ \mathcal{S} $ containing all the 128 products of generators (including the identity $ I $), whose elements are called \textit{stabilizers}.}
	\label{Stabilizer_state}
\end{figure}

The set $ \mathcal{S} $ containing all generators and their products (including the identity operator $ I $) forms the stabilizer group, containing $ 2^{N} $ elements, called stabilizers (Fig.~\ref{Stabilizer_state}(c)). We say that $ S $ spans $ \mathcal{S} $ and we will use the notation $ \left[  S\right]   =\mathcal{S} $. 

We will use the fact that the choice of the generator set is not unique, because generators contained in it can be recombined and still generate the same stabilizer state, but may require different measurement settings. A recombined generator set $ S^{(R)} $ contains products of the generators in $ S $, and these products are controlled by a non-singular $ N\times N $ binary matrix $ R $:
\begin{equation}\label{recombination_matrix}
S^{(R)}=\left\lbrace s_{i}^{(R)}:\;\;s_{i}^{(R)}=\prod_{j=1}^{N}s_{j}^{R_{ij}}\;\;,\;\;i=1,\ldots,N\right\rbrace	,
\end{equation}
where $ s_{j}^{0}=I $ and $ s_{j}^{1}=s_{j} $.

For instance, the following recombined generator set spans the same stabilizer as the generator set in Fig.~\ref{Stabilizer_state}(b):
\begin{equation}\label{alt_basis}
S^{(R)}=\left\lbrace s_{R}^{Z}s_{R}^{X}\,,s_{B}^{Z}\,s_{B}^{X}\,,\,s_{G}^{Z}s_{G}^{X}\,,\,s_{R}^{X}\,,\,s_{B}^{X}\,,\,s_{G}^{X}\,,\,s_{L}^{X}\right\rbrace	,
\end{equation}
where the $ Z $-type generators have been multiplied by the $ X $-type generators of the same plaquette, so they have transformed into stabilizers containing only $ Y $ Pauli matrices.

Recombinations preserve the stabilizer $ \mathcal{S} $, so they also preserve the projector onto the stabilizer state $ \ket{\mathcal{S}} $:
\begin{equation}\label{recombinations_preserve_projector}
\ket{\mathcal{S}}\bra{\mathcal{S}}=\prod_{s_{i}^{(R)}\in S^{(R)}}\frac{I+s_{i}^{(R)}}{2}=\frac{1}{2^{N}}\sum_{s_{i}\in\mathcal{S}}s_{i}\;\;\;\;\forall\;\;R	.
\end{equation}

Analogously, a \textit{generator subset} $ W_{\varOmega}\subset\mathcal{S} $ of $ n<N $ independent and commuting stabilizers of $ \mathcal{S} $ can be recombined by $ n\times n $ non-singular binary matrices $ R_{\varOmega} $. All the recombined generator subsets $ W_{\varOmega}^{(R_{\varOmega})} $ span the same subset of $ 2^{n} $ stabilizers (including the identity $ I $):
\begin{equation}
\left[  W_{\varOmega}^{(R_{\varOmega})}\right]  = \left[  W_{\varOmega}\right]\subset\mathcal{S}  \;\;\;\; \forall \;\; R_{\varOmega}	.
\end{equation}

Then, the projector onto the common $ +1 $ eigenspace of all the stabilizers in $ W_{\varOmega} $ is preserved under recombinations:
\begin{equation}
\begin{split}
\prod_{s_{i}^{(R_{\varOmega})}\in W_{\varOmega}^{(R_{\varOmega})}}\frac{I+s_{i}^{(R_{\varOmega})}}{2}&=\prod_{s_{j}\in W_{\varOmega}}\frac{I+s_{j}}{2}\\
&=\frac{1}{2^{n}}\sum_{s_{i}\in\left[  W_{\varOmega}\right]}s_{i}
\end{split}
\end{equation}
for all $ R_{\varOmega} $.

\subsection{Standard and modified genuine witnesses for stabilizer states}
For all non-separable multi-qubit stabilizer states $ \ket{\mathcal{S}} $ the maximal Schmidt coefficient of all bipartitions is $ \alpha_{\ket{\mathcal{S}}}=1/2 $, and the fidelity operator is the projector onto the $ +1 $ eigenvalue of the $ N $ stabilizers. Then, the \textbf{standard genuine witness} constructed for the stabilizer state reads
\begin{equation}\label{d}
\mathcal{W}=\frac{1}{2}\,I-\prod_{s_{i}\in S}\frac{I+s_{i}}{2}	.
\end{equation}

The standard genuine witness is the same if we use any recombined generator set $ S^{(R)} $ instead of $ S $ because, as pointed out above, recombinations preserve the projector of Eq.~(\ref{recombinations_preserve_projector}). For instance, the genuine witness for the seven-qubit color code is preserved if one uses the recombined generator set $ S^{(R)} $ in Eq.~(\ref{alt_basis}) instead of $ S $ in Fig.~\ref{Stabilizer_state}(b). To compute the expectation value of this witness it is necessary to measure $ 2^{N}-1 $ stabilizers, which becomes in general impractical for large $N$. One way to overcome this problem is to construct modified witness operators that includes a smaller number of terms. One option is to consider what we call the the \textbf{\textit{alternative witness}} \cite{Toth2005_94,Toth2005_72} in the following, 
\begin{equation}\label{g}
\mathcal{W}_{a}=\frac{N-1}{2}\,I-\frac{1}{2}\sum_{s_{i}\in S}s_{i}.
\end{equation}
This requires measuring only the $N$ stabilizers in a generator set $ S $. In this case, substituting $ S $ by a recombined generator set $ S^{(R)} $ changes the alternative witness, which however remains a genuine witness.

If the stabilizer state $ \ket{\mathcal{S}} $ has a generator set $ S $ composed entirely by $X$-type and $Z$-type stabilizers $ s^{X}_{i} $ and $ s^{Z}_{j} $ respectively, i.e.~each involving single-qubit Pauli operators $X$ and $Z$ only, (for instance Calderbank-Steane-Shor (CSS) codes \cite{Calderbank1996, Steane1996}), this allows to construct a \textbf{\textit{two-measurements witness}}, 
\begin{equation}\label{ea}
\mathcal{W}_{2m}=\frac{3}{2}\,I-\prod_{s^{X}_{i}\in S}\frac{I+s^{X}_{i}}{2} -\prod_{s^{Z}_{i}\in S}\frac{I+s^{Z}_{i}}{2}.
\end{equation}
Note that its evaluation requires only two measurement settings, with all qubits measured in the $X$ and the $Z$ basis.

For the two-measurement witness we have chosen the measurement settings $ X_{1}X_{2}\cdots X_{N} $ and $ Z_{1}Z_{2}\cdots Z_{N} $ but rotated versions $ \sigma_{1}\sigma_{2}\cdots \sigma_{N} $ and $ \sigma_{1}'\sigma_{2}'\cdots \sigma_{N}' $ can be used instead as long as $ \sigma_{\mu}\neq\sigma_{\mu}' $. For instance, in Ref.~\cite{Alba2010} the authors construct modified witnesses for bi-colorable graph states, where one measurement setting consists in measuring $X$ on every even qubit and $Z$ on every odd qubit, and vice versa for the second measurement setting. These operators are witnesses because their expectation value is larger than the expectation value of the standard genuine witness for any state \cite{Toth2005_94, Toth2005_72}, 
\begin{equation}\label{genuine_finnes}
\braket{\mathcal{W}}_{\rho}\leq\braket{\mathcal{W}_{a}}_{\rho}\;\;,\;\;\braket{\mathcal{W}}_{\rho}\leq\braket{\mathcal{W}_{2m}}_{\rho}\;\;\forall\,\rho	,
\end{equation}
which guarantees non-negative expectation values for any separable state $ \rho $.

The reduction in the number of measurement settings comes at the price that modified witnesses are in general less tolerant to noise. They detect entanglement only in a subset of the states where the standard genuine witness also detects entanglement.

\section{Local entanglement witness operators}\label{S3}
As stated above, we call a local witness an observable that detects the entanglement of a subsystem of qubits (Fig.~\ref{Local_entanglement}(a)). This type of entanglement coincides with localizable entanglement \cite{Verstraete2004,Popp2005,Sadhukhan2015}. In previous works, local witnesses were proposed for graph states in \cite{Alba2010,Nicolai2018}. Here, we discuss how they can be constructed and why they are detectors of entanglement. Then, we use the fact that every stabilizer state is equivalent to some graph state up to a local unitary (i.e.~a product of single-qubit unitary operations) \cite{Van2004} to show that local witnesses for stabilizer states also detect this entanglement. 

A local witness can be decomposed into a sum of stabilizers, and the number of stabilizers that must be measured to evaluate its expectation value only depends on the number of qubits in the subsystem. On the other hand, the evaluation of multiple local witnesses allows one to resolve the entanglement structure of a state. This requires the measurement of a number of stabilizers that grows with the number of subsystems in addition to the number of qubits in each one. However, using the same ideas applied to the modified genuine witnesses, it is possible to reduce the number of measurements. We show that in fact, only two measurement settings are often enough to reveal the entanglement structure with a modified version of the local witnesses.

\subsection{Graph states}
\begin{figure}[t]
	\centering
	\includegraphics[width=1\columnwidth]{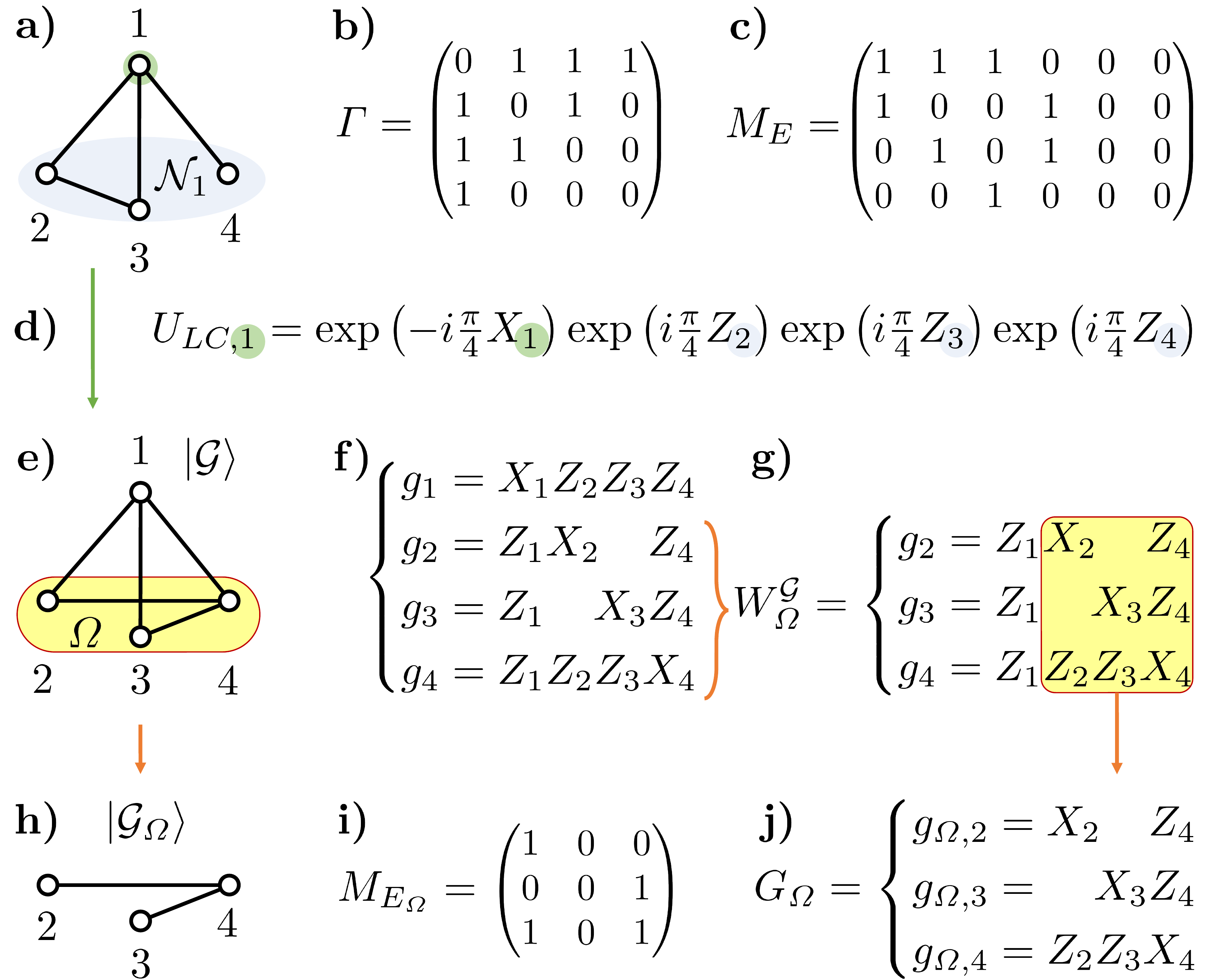}
	\caption{\textbf{Graph state concepts.} \textbf{(a)} Underlying graph of a 4-qubit graph state, and neighborhood $ \mathcal{N}_{1} $ of the site $ 1 $. \textbf{(b)} Adjacency matrix $ \varGamma $. \textbf{(c)} Incidence matrix $ M_{E} $. \textbf{(d)} Unitary operation $ U_{LC,1} $ that realizes a local complementation on the qubit $ 1 $. \textbf{(e)} Graph defining the locally equivalent graph state $ \ket{\mathcal{G}} $ obtained after applying $ U_{LC,1} $, and subsystem of qubits $ \varOmega=\{2,3,4\} $. \textbf{(f)} Natural generator set of a graph state. \textbf{(g)} Generator subset $ W_{\varOmega}^{\mathcal{G}} $ that constructs a local witness for $ \ket{\mathcal{G}} $ in $ \varOmega $. \textbf{(h)} Subgraph of the reduced graph state $ \ket{\mathcal{G_{\varOmega}}} $ in $ \varOmega $. \textbf{(i)} Reduced incidence matrix $ M_{E_{\varOmega}} $. The order of columns in $ M_{E_{\varOmega}} $ is not important for this work. \textbf{(j)} Basis of $ \ket{\mathcal{G_{\varOmega}}} $ composed by the reduced graph state generators $ g_{\varOmega,\mu} $.}
	\label{Graphs}
\end{figure}

We now review some key concepts of graph states that enable the construction of the local witnesses as proposed in \cite{Alba2010}. A \textbf{graph state} $ \ket{\mathcal{G}} $ is an $ N $-qubit stabilizer state \cite{Hein2006} associated to an underlying undirected graph, like the one depicted in Fig.~\ref{Graphs}(a), formed by $ N $ sites (representing the qubits) and a set $ E $ of links or edges $ l=\{\mu,\nu\} $ connecting sites $\mu$ and $\nu$. A graph, like the one in Fig.~\ref{Graphs}(a), can be represented by a $ N\times N $ adjacency matrix $ \varGamma $ with elements $ \varGamma_{\mu\nu}=1\,(0) $ if the sites $ \mu,\,\nu $ are (ar not) linked, 
or, equivalently, by the $ N\times{N \choose 2} $ \textit{incidence matrix} $ M_{E} $, 
\begin{equation}
\left[ M_{E}\right] _{\mu l}=\left\lbrace \begin{split}
&1 \;\;\mathrm{if} \;\;\mu\in l \;\;\mathrm{and}\;\;l\in E\\
&0 \;\;\mathrm{otherwise}
\end{split}\right.  ,
\end{equation}
with the index $ \mu $ indicating the $\mu$th site and $ l $ one of the $ {N \choose 2} $ pairs of sites.

A graph state is non-separable if and only if the underlying graph is connected  \cite{Hein2006}. A graph is \textit{connected} if it can not be separated into two components without breaking a link. The rank modulo 2 of $ M_{E} $ is related to the number $ m $ of connected components of the graph: $ \mathrm{rank}\left( M_{E}\right) =N-m $ \cite{West2000}. In particular, $\mathrm{rank}\left( M_{E}\right) =N-1 $, if the graph is connected.

For a graph state $ \ket{\mathcal{G}} $, there is one \textit{graph state generator} $ g_{\mu} $ associated to each qubit $\mu$, 
\begin{equation}\label{af}
g_{\mu}=X_{\mu}\prod_{\nu\in\mathcal{N}_{\mu}}Z_{\nu},
\end{equation}
where $ \mathcal{N}_{\mu} $  is the \textit{neighborhood} of the site $\mu$.

An equivalent representation of a graph state is 
\begin{equation}\label{ae}
\ket{\mathcal{G}}=\left[ \prod_{\left\lbrace \mu,\nu\right\rbrace \in E}U^{CZ}_{ \mu\nu}\right]  \bigotimes_{\eta=1}^{N}\frac{1}{\sqrt{2}}\left( \ket{0_{\eta}}+\ket{1_{\eta}}\right)	, 
\end{equation}
where controlled-phase gates 
\begin{equation}\label{controlled-phase}
U^{CZ}_{ \mu\nu}=\frac{I_{\nu}+Z_{\nu}}{2}I_{\mu}+\frac{I_{\nu}-Z_{\nu}}{2}Z_{\mu}
\end{equation}
act on all pair of qubits (sites) connected by a link in the underlying graph.
\\

We need to define analogous concepts within the subsystem $ \varOmega $ where we aim to detect entanglement. From a graph state it is possible to extract a \textit{reduced graph state} $ \ket{\mathcal{G}_{\varOmega}} $ for each subsystem $ \varOmega $ of $ n<N $ qubits. The underlying graph is the \textit{reduced graph} (Fig.~\ref{Graphs}(h)) composed by the sites in $ \varOmega $ and the subset of links $ E_{\varOmega}\subset E $ connecting them. It is represented by the \textit{reduced incidence matrix} $ M_{E_{\varOmega}} $ (Fig.~\ref{Graphs}(i)) that describes the links in $ E_{\varOmega} $.  Thus, the reduced graph is connected and hence, $ \ket{\mathcal{G}_{\varOmega}} $ is non-separable, if and only if:
\begin{equation}\label{reduced_inc_rank}
\mathrm{rank}\left( M_{E_{\varOmega}}\right) =n-1 .
\end{equation}

The graph state generators $ g_{\varOmega,\mu} $ of the reduced graph state $ \ket{\mathcal{G}_{\varOmega}} $ can be obtained from the generator subset of $ n $ graph state generators corresponding to the qubits in $ \varOmega $ (Fig.~\ref{Graphs}(g)):
\begin{equation}
W_{\varOmega}^{\mathcal{G}}=\left\lbrace g_{\mu}:\;\;\mu\in\varOmega\;\;,\;\;g_{\mu}\ket{\mathcal{G}}=\ket{\mathcal{G}}\right\rbrace	.
\end{equation}
From this set one can construct the \textit{reduced graph state generators} (Fig.~\ref{Graphs}(j)), 
\begin{equation}\label{reduced_graph_generators}
	g_{\varOmega,\mu}=\bigotimes_{\nu\in\varOmega} g_{\mu\nu} \;\;\;\;\forall\;\;\mu\in\varOmega.
\end{equation}
Here, $ g_{\mu\nu}\in\left\lbrace I,X,Z\right\rbrace $ is the single-qubit Pauli operator that appears for qubit $ \nu$ in the graph state generator $ g_{\mu}\in W_{\varOmega}^{\mathcal{G}} $.

These reduced graph state generators form a complete generator set $ G_{\varOmega} $ of the reduced graph state $ \ket{\mathcal{G}_{\varOmega}} $ (Fig.~\ref{Graphs}(i)),
\begin{equation}
g_{\varOmega,\mu}\ket{\mathcal{G}_{\varOmega}}=\ket{\mathcal{G}_{\varOmega}}\;\;\;\;\forall\;\;\mu\in\varOmega.
\end{equation}

Since the controlled-phase gates equal their inverse operation, the unitary that disentangles $ \varOmega $ from the rest $ \bar{\varOmega} $ of qubits $ U_{\mathrm{dis}}^{\mathcal{G}}\ket{\mathcal{G}}=\ket{\mathcal{G}_{\varOmega}}\otimes\ket{\mathcal{G}_{\bar{\varOmega}}} $ is the product of controlled-phase operations that break the links connecting $ \varOmega $ with $ \bar{\varOmega} $,
\begin{equation}\label{product_controlled}
U_{\mathrm{dis}}^{\mathcal{G}}=\prod_{\{\mu,\nu\}\in T}U^{CZ}_{\mu\nu}	.
\end{equation}
where
\begin{equation}
	T=\left\lbrace \{\mu,\nu\}\in E:\;\;\mu\in\varOmega\;,\;\nu\in\bar{\varOmega}\right\rbrace
\end{equation}

This unitary applied to the generator subset $ W_{\varOmega}^{\mathcal{G}} $ leads to the reduced graph state generators $ g_{\varOmega,\mu} $ (Fig.~\ref{Graphs}(j))
\begin{equation}
	U_{\mathrm{dis}}^{\mathcal{G}}g_{\mu}U_{\mathrm{dis}}^{\mathcal{G}\dagger}=g_{\varOmega,\mu}\otimes I_{\bar{\varOmega}}, 
\end{equation}
and hence
\begin{equation}\label{projector}
U_{\mathrm{dis}}^{\mathcal{G}}\left[ \prod_{g_{\mu}\in W_{\varOmega}^{\mathcal{G}}}\frac{I+g_{\mu}}{2}\right] U_{\mathrm{dis}}^{\mathcal{G}\dagger}=\ket{\mathcal{G}_{\varOmega}}\bra{\mathcal{G}_{\varOmega}}\otimes I_{\bar{\varOmega}}.
\end{equation}

\subsection{Local witnesses for graph states}\label{graph_witnesses}
Following the approach in \cite{Alba2010}, the entanglement of an $ N $-qubit state $ \rho $ is localized in a subsystem $\varOmega$ by means of local measurements on the rest of qubits $\bar{\varOmega}$. The entanglement does not increase under these measurements, so the remaining entanglement in the reduced state $ \rho_{\varOmega} $ obtained after the measurement was present in $\rho$. On $ \rho_{\varOmega} $ a genuine witness $ \mathcal{W} $ is evaluated to detect the remaining entanglement. The key observation is that, instead of executing these measurements, we can instead evaluate a local entanglement witness $ \mathcal{W}_{\varOmega} $ on the global state $ \rho $ because the expectation values coincide, $ \braket{\mathcal{W}}_{\rho_{\varOmega}}=\braket{\mathcal{W}_{\varOmega}}_{\rho} $. Therefore, a negative expectation value of the local witness $ \mathcal{W}_{\varOmega} $ detects the presence of entanglement in the reduced state $ \rho_{\varOmega} $, and thus the entanglement in the subsystem $ \varOmega $ of the global state $ \rho $ is detected.
\\

Let us briefly recall the effect of $ Z $ measurements on a pure graph state $ \ket{\mathcal{G}} $. All $2^{N-n} $ possible outcomes of the measurements of qubits $\nu$ in $ \bar{\varOmega} $ can be represented by a binary number $x$ of length $N-n$, with the bit $ x_{\nu} = 0\, (1)$ for outcome $ +1\,(-1)$. The reduced state obtained from $\ket{G}$ after the outcome $x$ on the measurement of $ Z $ on every qubit $ \nu\in\bar{\varOmega} $ is $ \boldsymbol{Z}_{x}\ket{\mathcal{G}_{\varOmega}} $ \cite{Hein2006}, i.e. the reduced graph state $ \ket{\mathcal{G}_{\varOmega}} $ up to outcome-dependent single-qubit $Z$ operations, $ Z_{\mu}^{0}=I_{\mu} $ and $ Z_{\mu}^{1}=Z_{\mu} $. The effect of the latter is equivalent to the action of the disentangling unitary in Eq.~(\ref{product_controlled}) and tracing over the rest of qubits, 
\begin{equation}
	\ket{\mathcal{G}_{\varOmega}}\bra{\mathcal{G}_{\varOmega}}=\mathrm{tr}_{\bar{\varOmega}}\left( U_{\mathrm{dis}}^{\mathcal{G}}\ket{\mathcal{G}}\bra{\mathcal{G}}U_{\mathrm{dis}}^{\mathcal{G}\dagger}\right).
\end{equation}
If the same process is applied to a mixed $ N $-qubit state $ \rho $, the reduced state is
\begin{equation}
	\rho_{\varOmega}^{\mathcal{G}}=\mathrm{tr}_{\bar{\varOmega}}\left( U_{\mathrm{dis}}^{\mathcal{G}}\rho U_{\mathrm{dis}}^{\mathcal{G}\dagger}\right).
\end{equation}
One can check that this reduced state is the average over all possible outcomes states $ \rho_{\varOmega,x} $ obtained after the same measurement process, multiplied by the outcome-dependent single-qubit operations, 
\begin{equation}
	\rho_{\varOmega}^{\mathcal{G}}=\sum_{x}p_{x}\boldsymbol{Z}_{x}\rho_{\varOmega,x}\boldsymbol{Z}_{x}^{\dagger}	.
\end{equation}
The normalized state corresponding to outcome $ x $
\begin{equation}
	\rho_{\varOmega,x}=\frac{1}{p_{x}}\mathrm{tr}_{\bar{\varOmega}}\left(\rho\prod_{\nu\in\bar{\varOmega}}\frac{I_{\nu}+(-1)^{x_{\nu}}Z_{\nu}}{2}\right)
\end{equation}
is obtained with a probability $ p_x $.
\\
Once the reduced state $\rho_{\varOmega}^{\mathcal{G}}$ is obtained, a genuine witness of the form Eq.~(\ref{a})
\begin{equation}
	\mathcal{W}^{\mathcal{G}_{\varOmega}}=\alpha_{\ket{\mathcal{G}_{\varOmega}}}I_{\varOmega}-\ket{\mathcal{G}_{\varOmega}}\bra{\mathcal{G}_{\varOmega}}
\end{equation}
allows one to detect entanglement in the region. Again, the expectation value of this genuine witness evaluated in $\rho_{\varOmega}^{\mathcal{G}}$ equals the expectation value of the following local witness evaluated in the original state $\rho$, 
\begin{equation}\label{local_witness}
	\mathcal{W}_{\varOmega}^{\mathcal{G}}=\alpha_{\ket{\mathcal{G}_{\varOmega}}}I-\prod_{g_{\mu}\in W_{\varOmega}^{\mathcal{G}}}\frac{I+g_{\mu}}{2}	,
\end{equation}
yielding, using Eq.~(\ref{projector}), 
\begin{equation}
	\braket{\mathcal{W}^{\mathcal{G}_{\varOmega}}}_{\rho_{\varOmega}^{\mathcal{G}}}=\braket{\mathcal{W}_{\varOmega}^{\mathcal{G}}}_{\rho}	.
\end{equation}

As pointed out in section \ref{genuine_detectors}, $\mathcal{W}^{\mathcal{G}_{\varOmega}}$ acts as a witness only if $\ket{\mathcal{G}_{\varOmega}}$ is a non-separable state. Therefore, $W_{\varOmega}^{\mathcal{G}}$ must be chosen such that the genuine witness $ \mathcal{W}^{\mathcal{G}_{\varOmega}} $ is prepared for a non-separable graph state $\ket{\mathcal{G}_{\varOmega}}$.

This expectation value is the average of the genuine witness evaluated on all reduced states after the measurement, 
\begin{equation}
	\braket{\mathcal{W}_{\varOmega}^{\mathcal{G}}}_{\rho}=\sum_{x}p_{x}\braket{\mathcal{W}^{\mathcal{G}_{\varOmega}}}_{\boldsymbol{Z}_{x}\rho_{\varOmega,x}\boldsymbol{Z}_{x}^{\dagger}}	.
\end{equation}

Since $ p_{x} $ are positive coefficients, if the expectation value of the local witness is negative, at least one $ \braket{\mathcal{W}^{\mathcal{G}_{\varOmega}}}_{\boldsymbol{Z}_{x}\rho_{\varOmega,x}\boldsymbol{Z}_{x}^{\dagger}} $ with $ p_{x}\neq0 $ is also negative, indicating that an outcome state $ \rho_{\varOmega,x} $ is non-separable (recall that $ \boldsymbol{Z}_{x} $ preserves entanglement). Thus, the entanglement has been localized in the subsystem $ \varOmega $ by the corresponding measurement.

This is due to the connection between local witnesses and \textit{localizable entanglement} \cite{Verstraete2004,Popp2005,Sadhukhan2015}. This entanglement measure provides the maximum amount of entanglement $ L_{\varOmega}(\rho) $ that can be localized on average in a subsystem $ \varOmega $ by means of local measurements on the rest of qubits. Every measurement $ \mathcal{M} $ specifies a set of $ 2^{N-n} $ outcome states $ \rho_{\varOmega,x} $ and probabilities $ p_{x} $, so for a given $ \mathcal{M} $ the average entanglement localized is:
\begin{equation}
L_{\varOmega}(\rho)=\max_{\mathcal{M}}L_{\varOmega}^{\mathcal{M}}(\rho)\;\;,\;\;L_{\varOmega}^{\mathcal{M}}(\rho)=\sum_{x}p_{x}E\left( \rho_{\varOmega,x}\right)	,	
\end{equation}
where $ E\left( \rho_{\varOmega,x}\right)  $ is an entanglement measure.

As we explained, if a local witness has negative expectation value, there is genuine entanglement in one of the outcome states $ \rho_{\varOmega,x} $ with $ p_{x}\neq0 $ resulting of the particular measurement of $ Z $. Hence, the amount $ E\left( \rho_{\varOmega,x}\right)\neq0  $ is non-zero and consequently, the localizable entanglement is non-zero $ L_{\varOmega}(\rho)\neq0 $. We can conclude that local witnesses detect the localizable entanglement because a negative expectation value of the first implies that the value of the second is non-zero.

\subsection{Local witness operators for stabilizer states}\label{S2s1}
Now we want to construct a witness for the stabilizer state $ \ket{\mathcal{S}} $, i.e.~we want to measure only stabilizers $ s_{i}\in\mathcal{S} $ of its stabilizer. We can show that if we find a generator subset $ W_{\varOmega}\subset\mathcal{S} $ that is locally equivalent to some generator subset of graph state generators $ W_{\varOmega}^{\mathcal{G}} $, the local witness constructed with $ W_{\varOmega} $ instead of $ W_{\varOmega}^{\mathcal{G}} $ is also a local witness that detects the entanglement of $ \varOmega $.

We say that one generator subset $ W_{\varOmega} $ of stabilizers is \textit{locally equivalent} (LE) to a generator subset $ W_{\varOmega}^{\mathcal{G}} $ of graph state generators if there is a local unitary $ V_{LE} $ and a recombination $ R_{\varOmega} $ such that:
\begin{equation}
	V_{LE}W_{\varOmega}^{(R_{\varOmega})}V_{LE}^{\dagger}=W_{\varOmega}^{\mathcal{G}}	.
\end{equation}

Hence, we can construct the following local witness that is decomposed only into stabilizers of the stabilizer state $ \ket{\mathcal{S}} $ and we will call \textit{\textbf{standard local witness}}:
\begin{equation}\label{w}
\mathcal{W}_{\varOmega}=\frac{1}{2}\,I-\prod_{s_{i}\in W_{\varOmega}}\frac{I+s_{i}}{2}	.
\end{equation}
Note that the standard local witness is locally equivalent to the witness for a graph state $ V_{LE}\mathcal{W}_{\varOmega}V_{LE}^{\dagger}=\mathcal{W}_{\varOmega}^{\mathcal{G}} $. Then, the expectation value of $ \mathcal{W}_{\varOmega} $ evaluated on an state $ \rho $ coincides with the expectation value of $ \mathcal{W}_{\varOmega}^{\mathcal{G}} $ evaluated on the locally equivalent state $ V_{LE}\rho V_{LE}^{\dagger} $, 
\begin{equation}
	\braket{\mathcal{W}_{\varOmega}}_{\rho}=\braket{\mathcal{W}_{\varOmega}^{\mathcal{G}}}_{V_{LE}\rho V_{LE}^{\dagger}},
\end{equation}
which has the same entanglement properties. Again, $ \mathcal{W}_{\varOmega}^{\mathcal{G}} $ is a witness only if the reduced graph state $ \ket{\mathcal{G}_{\varOmega}} $ associated to $ W_{\varOmega}^{\mathcal{G}} $ is non-separable.
\\

If the standard local witness has negative expectation value, the entanglement detected can be understood in an analogous way. In this case, the measurements of the qubit $ \nu\in\bar{\varOmega} $ outside the subsystem are performed in the basis $ V_{LE}^{-1}Z_{\nu}V_{LE} $. The outcome-dependent correction is $ V_{LE}^{-1}\boldsymbol{Z}_{x}V_{LE} $, which is again a local unitary operation that does not change the entanglement properties of the outcome states. The average of all the outcome states corrected in this way is
\begin{equation}
	\rho_{\varOmega}=\mathrm{tr}_{\bar{\varOmega}}\left( U_{\mathrm{dis}}\rho U_{\mathrm{dis}}^{\dagger}\right).	
\end{equation}
Here the disentangling unitary is
\begin{equation}
	U_{\mathrm{dis}}=V_{LE}^{-1}U_{\mathrm{dis}}^{\mathcal{G}}V_{LE}	.
\end{equation}

Then, a negative expectation value of the local witness implies that at least one outcome state is genuinely entangled, and therefore entanglement has been localized in $ \varOmega $.\\

Now, the problem reduces to finding the generator subset $ W_{\varOmega} $ of stabilizers that constructs a valid local witness for the subsystem $ \varOmega $. The conditions on $ W_{\varOmega} $ are summarized in the next Proposition.

\begin{proposition}
	The stabilizer state $ \ket{\mathcal{S}} $ with stabilizer $ \mathcal{S} $ has a local witness of the form of Eq.~(\ref{w}) for the subsystem $ \varOmega $ that is constructed from the generator subset $ W_{\varOmega}\subset\mathcal{S} $ of stabilizers if and only if the following conditions are satisfied:
	\begin{enumerate}[label=(\Alph*)]	
		\item $ W_{\varOmega} $ is locally equivalent to some generator subset of graph state generators, i.e.~there is a local unitary $ V_{LE} $ and a recombination $ R_{\varOmega} $ such that:
		\begin{equation}\label{subset_equivalence}
		V_{LE}W_{\varOmega}^{(R_{\varOmega})}V_{LE}^{\dagger}=W_{\varOmega}^{\mathcal{G}}	.
		\end{equation}
		\item The reduced graph state $ \ket{\mathcal{G}_{\varOmega}} $ is non-separable.
	\end{enumerate}
	\label{p}
\end{proposition}

This takes into account that it is possible that a generator subset $ W_{\varOmega} $ can not be transformed directly into some $ W_{\varOmega}^{\mathcal{G}} $ by a local unitary, but that there is a recombined generator subset $ W_{\varOmega}^{(R_{\varOmega})} $ for which such a local unitary exists. For instance, the following generator subsets of stabilizers of the seven-qubit color code differ only in a recombination $ R_{\varOmega} $ but the first one can not be transformed directly into graph state generators:
\begin{eqnarray}\label{e1}
	W_{\{1,2,3,4\}}&=\left\lbrace s_{R}^{X}s_{R}^{Z}\;,\;s_{R}^{Z}\;,\;s_{B}^{Z}\;,\;s_{G}^{Z}\right\rbrace\\\label{e2}
	W_{\{1,2,3,4\}}^{(R_{\varOmega})}&=\left\lbrace s_{R}^{X}\;,\;s_{R}^{Z}s_{B}^{Z}s_{G}^{Z}\;,\;s_{B}^{Z}\;,\;s_{G}^{Z}\right\rbrace	.
\end{eqnarray}

More concretely, the first two elements $ s_{R}^{X}s_{R}^{Z}=Y_{1}Y_{2}Y_{3}Y_{4} $ and $ s_{R}^{Z}=Z_{1}Z_{2}Z_{3}Z_{4} $ of $ W_{\{1,2,3,4\}} $ can not be transformed into two graph state generators with the same local unitary $ V_{LE} $. But the local unitary $ V_{LE}=H_{1}H_{2}H_{4} $ composed by three Hadamard operators transforms the stabilizers in $ W_{\{1,2,3,4\}}^{(R_{\varOmega})} $ into graph state generators:
\begin{equation}
	\begin{split}
	s_{R}^{X}=X_{1}X_{2}X_{3}X_{4}&\;\;\mapsto\;\; g_{3}=Z_{1}Z_{2}X_{3}Z_{4}\\
	s_{R}^{Z}s_{B}^{Z}s_{G}^{Z}=Z_{1}Z_{3}Z_{5}Z_{7}&\;\;\mapsto\;\; g_{1}=X_{1}Z_{3}Z_{5}Z_{7}\\
	s_{B}^{Z}=Z_{2}Z_{3}Z_{5}Z_{6}&\;\;\mapsto\;\; g_{2}=X_{2}Z_{3}Z_{5}Z_{6}\\
	s_{G}^{Z}=Z_{3}Z_{4}Z_{6}Z_{7}&\;\;\mapsto\;\; g_{4}=Z_{3}X_{4}Z_{6}Z_{7}
	\end{split}	.
	\label{mapping}
\end{equation}

Note also that the only condition of the local unitary $ V_{LE} $ is that it transforms a generator subset $ W_{\varOmega} $ of the stabilizer state $ \ket{\mathcal{S}} $ into a generator subset $ W_{\varOmega}^{\mathcal{G}} $ of a graph state $ \ket{\mathcal{G}} $, but it is not forced to transform $ \ket{\mathcal{S}} $ into $ \ket{\mathcal{G}} $. This fact will become important below where we prove that the direct method indeed finds more witnesses than the graph-based method.

\subsection{Modified local witnesses}
The evaluation of the standard local witness of Eq.~(\ref{w}) requires the measurement of $ 2^{n}-1 $ stabilizers. This number can be reduced further following the same techniques employed for the modified genuine witnesses. Once a valid subset $ W_{\varOmega} $ that satisfies Proposition \ref{p} is found, it can be used to construct modified local witnesses that require less measurement settings. For instance, with $ n $ measurements we can compute the value of the \textit{\textbf{alternative local witness}}:
\begin{equation}\label{la}
\mathcal{W}_{\varOmega,a}=\frac{n-1}{2}\,I-\frac{1}{2}\sum_{s_{i}\in W_{\varOmega}}s_{i}	.
\end{equation}

\begin{table*}[]
	\centering
	\renewcommand{\arraystretch}{2}	
	\begin{tabular}{ C{7em} || C{7em} | C{24em} C{6em} C{8em}}
		\multicolumn{2}{c|}{} & Formula & Meas. set. & Crit. prob. $ p_{c} $\\
		\hline \hline
		\multirow{3}{4em}{Genuine witnesses} & \cellcolor[HTML]{DCDEFD} Standard & \cellcolor[HTML]{DCDEFD} \(\displaystyle \mathcal{W}=\frac{1}{2}\,I-\prod_{s_{i}\in S}\frac{I+s_{i}}{2} \)   & \cellcolor[HTML]{DCDEFD} $ 2^{N}-1 $ &\cellcolor[HTML]{DCDEFD}  $ \simeq1/2 $ \\&
		\cellcolor[HTML]{F6F7FF} Alternative &  \cellcolor[HTML]{F6F7FF} \(\displaystyle \mathcal{W}_{a}=\frac{N-1}{2}\,I-\frac{1}{2}\sum_{s_{i}\in S}s_{i} \)     &\cellcolor[HTML]{F6F7FF}  $ N $   & \cellcolor[HTML]{F6F7FF} $ 1-1/N $           \\&
		\cellcolor[HTML]{DCDEFD} Two measurements   &  \cellcolor[HTML]{DCDEFD} \(\displaystyle \mathcal{W}_{2m}=\frac{3}{2}\,I-\prod_{s^{X}_{i}\in S}\frac{I+s^{X}_{i}}{2} -\prod_{s^{Z}_{i}\in S}\frac{I+s^{Z}_{i}}{2}  \) & \cellcolor[HTML]{DCDEFD} $ 2 $  & \cellcolor[HTML]{DCDEFD} $ \in[2/3,3/4] $              \\
		\hline
		
		\multirow{3}{4em}{Local witnesses} &\cellcolor[HTML]{F6F7FF}  Standard & \cellcolor[HTML]{F6F7FF} \(\displaystyle \mathcal{W}_{\varOmega}=\frac{1}{2}\,I-\prod_{s_{i}\in W_{\varOmega}}\frac{I+s_{i}}{2} \)  & \cellcolor[HTML]{F6F7FF} $ 2^{n}-1 $ & \cellcolor[HTML]{F6F7FF} $ \in[1/3,1/2] $ \\&
		\cellcolor[HTML]{DCDEFD} Alternative  & \cellcolor[HTML]{DCDEFD} \(\displaystyle \mathcal{W}_{\varOmega,a}=\frac{n-1}{2}\,I-\frac{1}{2}\sum_{s_{i}\in W_{\varOmega}}s_{i} \) &\cellcolor[HTML]{DCDEFD} $ n $  & \cellcolor[HTML]{DCDEFD}$ 1-1/n $ \\&
		\cellcolor[HTML]{F6F7FF} Two measurements &  \cellcolor[HTML]{F6F7FF} \(\displaystyle \mathcal{W}_{\varOmega,2m}=\frac{3}{2}\,I-\prod_{s_{i}^{X}\in W_{\varOmega}}\frac{I+s^{X}_{i}}{2} -\prod_{s_{i}^{Z}\in W_{\varOmega}}\frac{I+s^{Z}_{i}}{2}  \)    & \cellcolor[HTML]{F6F7FF} $ 2 $   & \cellcolor[HTML]{F6F7FF} $ \in[1/2,3/4] $
	\end{tabular}
	\caption{Overview of genuine and local entanglement witnesses: the table shows the explicit form in terms of stabilizer operators $ s_{i} $ defining the $ N $-qubit stabilizer state, the maximum number of measurement settings required, and the critical probability of the white noise state of Eq.~(\ref{h}) below which no entanglement is detected.}
	\label{t1}
\end{table*}

Similar to the two-measurement local witness, if the subset $ W_\varOmega $ is entirely composed by $ X $-type and $ Z $-type stabilizers, one can construct a \textbf{\textit{two-measurements local witness}}:
\begin{equation}\label{le}
\mathcal{W}_{\varOmega,2m}=\frac{3}{2}\,I-\prod_{s_{i}^{X}\in W_{\varOmega}}\frac{I+s^{X}_{i}}{2} -\prod_{s_{i}^{Z}\in W_{\varOmega}}\frac{I+s^{Z}_{i}}{2}	,
\end{equation}
that can be evaluated with just two measurement settings. If $ W_{\varOmega} $ is not of this form, a recombined generator subset $ W_{\varOmega}^{(R_{\varOmega})} $ can be used instead. For instance, the generator subset in Eq.~(\ref{e1}) is not of this form because the first element contains Pauli operators $ Y $, but the generator subset in Eq.~(\ref{e2}) has the right form. In this example one can construct the following two-measurements local witness from the generator subset $ W_{\{1,2,3,4\}}^{(R_{\varOmega})} $:
\begin{equation}
\begin{split}
	\mathcal{W}_{\{1,2,3,4\},2m}&=\frac{3}{2}\,I-\frac{I+s^{X}_{R}}{2}\\
	&-\frac{I+s_{R}^{Z}s_{B}^{Z}s_{G}^{Z}}{2}\frac{I+s_{B}^{Z}}{2}\frac{I+s_{G}^{Z}}{2}
\end{split}	.
\end{equation}
Again, here we have chosen the measurement settings $X_{1}X_{2}\cdots X_{N}$ and $Z_{1}Z_{2}\cdots Z_{N}$, but other pairs can be chosen as well.

Like in Eq.~(\ref{genuine_finnes}), the modified local witnesses are valid witnesses as their expectation value is greater or equal than the expectation value of the standard local witness for every state, 
\begin{equation}\label{finnermodified}
\braket{\mathcal{W}_{\varOmega}}_{\rho}\leq\braket{\mathcal{W}_{\varOmega,a}}_{\rho}\;\;,\;\;\braket{\mathcal{W}_{\varOmega}}_{\rho}\leq\braket{\mathcal{W}_{\varOmega,2m}}_{\rho}\;\;\forall\,\rho.
\end{equation}
Again, the price of reducing the number of measurements is that the modified local witnesses are in general less tolerant to noise than standard local witnesses.

\subsection{Robustness and comparison of genuine and local witnesses}
To obtain a benchmark figure of merit for the robustness of a local witness, we consider \cite{Toth2005_72} the performance of the witnesses with respect to a noisy model state of the form 
\begin{equation}\label{h}
\rho_{p}=p\ket{\mathcal{S}}\bra{\mathcal{S}}+\frac{1-p}{2^{N}}I, 
\end{equation}
which is a mixture of the ideal stabilizer state $ \ket{\mathcal{S}} $ and the completely mixed state.

The \textit{robustness} of a witness is then given by the critical probability $ p_{c} $, that guarantees that the expectation value of the witness operator evaluated in the state $ \rho_{p} $ is negative, and thus entanglement is detected. The critical probabilities of all witnesses discussed summarized in Table \ref{t1}.
\\

Moreover, one witness $ \mathcal{W} $ is said to be \textit{finer} than other $ \mathcal{W}' $ if the expectation value of $ \mathcal{W} $ is negative on every state where the expectation value of $ \mathcal{W}' $ is negative. Thus, there is a positive coefficient $ \beta>0 $ such that $ \braket{\mathcal{W}}_{\rho}\leq\beta\braket{\mathcal{W}'}_{\rho} $ for any state $ \rho $. In that case, one can be sure that $ \mathcal{W} $ detects entanglement if $ \mathcal{W}' $ detects it. If one witness is finer than another, it is also more robust. 

There is a hierarchy between the presented witnesses considering this criteria. First, the standard genuine witness is finer than the modified genuine witnesses \cite{Toth2005_94, Toth2005_72}:
\begin{equation}
\braket{\mathcal{W}}_{\rho}\leq\braket{\mathcal{W}_{a}}_{\rho}\;\;,\;\;\braket{\mathcal{W}}_{\rho}\leq\braket{\mathcal{W}_{2m}}_{\rho}\;\;\forall\,\rho	,
\end{equation}
which again shows that reducing the number of measurements by means of modifying the witnesses leads to a decrease of the tolerance to noise. Similarly, and using the same ideas as in \cite{Toth2005_94, Toth2005_72}, the standard local witness constructed with the generator subset $ W_{\varOmega} $ is finer than the modified witnesses constructed from the same generator subset:
\begin{equation}
\braket{\mathcal{W}_{\varOmega}}_{\rho}\leq\braket{\mathcal{W}_{\varOmega,a}}_{\rho}\;\;,\;\;\braket{\mathcal{W}_{\varOmega}}_{\rho}\leq\braket{\mathcal{W}_{\varOmega,2m}}_{\rho}\;\;\forall\,\rho	.
\end{equation}

Second, standard local witnesses, for every subsystem $ \varOmega $, are finer than the standard genuine witness because the expectation value of the projector in Eq.~(\ref{w}) is bigger than the expectation value of the projector in Eq.~(\ref{d}) for any state $ \rho $:
\begin{equation}\label{finnergenuine}
\braket{\mathcal{W}_{\varOmega}}_{\rho}\leq\braket{\mathcal{W}}_{\rho}\;\;\forall\,\rho	.
\end{equation}

Third, two different standard local witnesses $ \mathcal{W}_{\varOmega} $ and $ \mathcal{W}_{\varOmega'} $ such that the generator subsets $ W_{\varOmega} $ and $ W_{\varOmega'} $ satisfy $ \left[ W_{\varOmega}\right] \subset \left[ W_{\varOmega'}\right] $ (therefore, the subsystem $ \varOmega $ is included inside the second $ \varOmega\subset\varOmega' $) satisfy that the first one is finer than the second:
\begin{equation}\label{finer}
\braket{\mathcal{W}_{\varOmega}}_{\rho}\leq\braket{\mathcal{W}_{\varOmega'}}_{\rho}\;\;\forall\,\rho	.
\end{equation}

For instance, the standard local witness constructed from $ W_{\{2,3\}}=\left\lbrace s_{R}^{X},s_{B}^{Z}\right\rbrace  $ is finer than the standard local witness constructed from $ W_{\{2,3,4\}}=\left\lbrace s_{R}^{X},s_{B}^{Z},s_{G}^{Z}\right\rbrace  $.
\\

In the following we provide recipes how to construct the local witnesses starting from the stabilizers of the state of interest.

\section{Construction of local witnesses}\label{S4}
We now propose two methods to find a generator subset $ W_{\varOmega} $ that satisfies Proposition \ref{p}: The \textit{\textbf{graph-based method}} exploits the connection between stabilizers and graph states, and the \textit{\textbf{direct method}} provides sufficient and necessary criteria to decide whether an arbitrary $ W_{\varOmega} $ does satisfy this proposition. The latter does not require finding the local unitary and the graph state generators (Figs.~\ref{Local_entanglement}(e.1) and \ref{Local_entanglement}(e.2)). We anticipate that the direct method finds the complete set of local stabilizer-based witnesses, while the graph-based method in general only finds a subset of them.

Moreover, we will show that generally there are multiple local witnesses for every subsystem. These witnesses act like different detectors of the same entanglement in the sense that the positive detection of just one of them is enough to confirm that the subsystem is entangled. 

\subsection{Graph-based method}\label{graph_method_one_graph}

\begin{figure}[t]
	\centering
	\includegraphics[width=1\columnwidth]{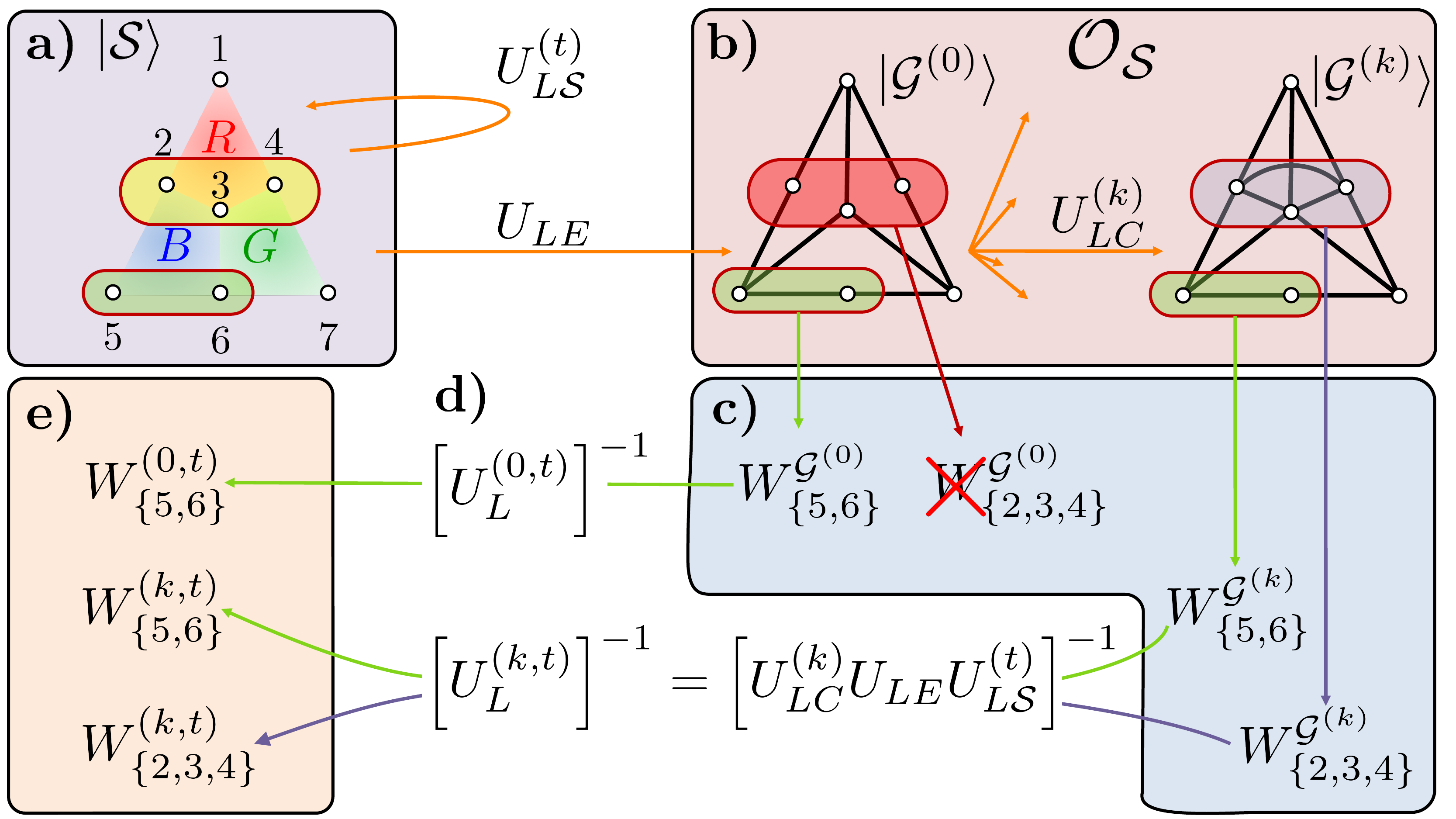}
	\caption{\textbf{Illustration of the graph-based method.} Here, local witnesses for \textbf{(a)} two subsystem $ \{5,6\} $ (green) and $ \{2,3,4\} $ (yellow) of the seven-qubit color code $ \ket{\mathcal{S}} $ are constructed. The protocol consists of the following steps: Find the local symmetries $ U_{L\mathcal{S}}^{(t)} $ such that $ U_{L\mathcal{S}}^{(t)}\ket{\mathcal{S}}=\ket{\mathcal{S}} $, and a local unitary $ U_{LE} $ that generates the initial locally equivalent graph state $ U_{LE}\ket{\mathcal{S}}=\ket{\mathcal{G}^{(0)}} $. \textbf{(b)} With local complementation unitary operations $ U_{LC}^{(k)} $ generate the orbit $ \mathcal{O}_{\mathcal{S}} $ of graph states locally equivalent to $ \ket{\mathcal{S}} $, i.e.~$ U_{LC}^{(k)}\ket{\mathcal{G}^{(0)}}=\ket{\mathcal{G}^{(k)}} $. For each graph consider the subsystems that are connected (e.g.~$ \{5,6\} $ is connected in both depicted graphs, but $ \{2,3,4\} $ is connected only in the second) and take the \textbf{(c)} generator subsets $ W_{\varOmega}^{\mathcal{G}^{(k)}} $. \textbf{(d)} Apply the local unitary $ \left[U_{L}^{(k,t)}\right]^{-1} $ to obtain \textbf{(e)} the generator subsets $ W_{\varOmega}^{(k,t)} $ that construct local witnesses for the stabilizer state $ \ket{\mathcal{S}} $.}
	\label{Graph_method}
\end{figure}

We now show how can one use the knowledge of a graph state $\ket{\mathcal{G}}$ that is locally equivalent to a stabilizer state of interest $\ket{\mathcal{S}}$, i.e.~$U_{LE}\ket{\mathcal{S}}=\ket{\mathcal{G}}$, to construct local witnesses for this state as illustrated in Fig.~\ref{Graph_method}. Some key properties of graph states are depicted in Fig.~\ref{Graphs}. It is known that for any stabilizer state such a locally equivalent graph state exists \cite{Van2004}. Within the binary picture of the stabilizer formalism, which we summarize in Appendix \ref{BIN_PIC}, the problem of finding such a graph state reduces to solving a system of binary equations \cite{Hein2006,Lang2012}.

Once the graph state generators have been found, the following stabilizer generator subset for state $ \ket{\mathcal{S}} $ in a region $\varOmega$ of interest can be obtained by transforming the graph state generators corresponding to qubits within  $\varOmega$, 
\begin{equation}
	W_{\varOmega}=\left\lbrace U_{LE}^{-1}g_{\mu}U_{LE}\in\mathcal{S}:\;\;\mu\in\varOmega\right\rbrace, 
\end{equation}
which satisfies property $ (A) $ in Proposition \ref{p}. The subset $W_{\varOmega}$ also satisfies property $ (B) $ if the underlying graph $ \varGamma $ of $ \ket{\mathcal{G}} $ is connected within the subsystem $ \varOmega $, and this is the case if the incidence matrix $ M_{E_{\varOmega}} $ of the reduced graph has rank $ n-1 $.

We illustrate this for the example of the first graph state in Fig.~\ref{Graph_method}(b). The local unitary  $ U_{LE}=H_{1}H_{5}H_{7} $ transforms the seven-qubit stabilizer color code as defined in Fig.~\ref{Stabilizer_state} into a graph state with the following correspondence between stabilizers and graph state generators:
\begin{equation}
\begin{split}
U_{LE}^{-1}\,g_{1}\,U_{LE}&= s_{R}^{Z}\;\;\;\;\;\;\;\;,\;\;\;\;\;\;\,\;U_{LE}^{-1}\,g_{2}\,U_{LE}= s_{G}^{X}s_{L}^{X}\\
U_{LE}^{-1}\,g_{3}\,U_{LE}&= s_{R}^{X}s_{B}^{X}s_{G}^{X}\;\;\;,\;\;\;U_{LE}^{-1}\,g_{4}\,U_{LE}= s_{B}^{X}s_{L}^{X}\\
U_{LE}^{-1}\,g_{5}\,U_{LE}&= s_{B}^{Z}\;\;\;\;\;\;\;\;,\;\;\;\;\;\,\;\;U_{LE}^{-1}\,g_{6}\,U_{LE}= s_{R}^{X}s_{L}^{X}\\
U_{LE}^{-1}\,g_{7}\,U_{LE}&= s_{G}^{Z}
\end{split}	.
\end{equation}

Whereas any generator subset $ W_{\varOmega} $ of these stabilizers satisfies property $ (A) $ in Proposition \ref{p}, this is not the case for property $ (B) $, which depends on the selected subsystem. For instance, since the subsystem $ \varOmega=\{5,6\} $ shown in green is connected, the generator subset $ W_{\{5,6\}}=\left\lbrace s_{B}^{Z}\,,\,s_{R}^{X}s_{L}^{X}\right\rbrace  $ indeed constructs a local witness operator, 
\begin{equation}
	\mathcal{W}_{\{5,6\}}=\frac{1}{2}\,I-\frac{I+s_{B}^{Z}}{2}\frac{I+s_{R}^{X}s_{L}^{X}}{2}
\end{equation}
for this subsystem. In contrast, the subsystem $ \{2,3,4\} $ in red is not connected within this graph, thus the generator subset $ W_{\{2,3,4\}}=\left\lbrace s_{G}^{X}s_{L}^{X}\,,\,s_{R}^{X}s_{B}^{X}s_{G}^{X}\,,\,s_{B}^{X}s_{L}^{X}\right\rbrace  $ does not provide a local witness. This illustrates that construction of a witness using the graph-based approach requires finding a locally equivalent graph state, for which the qubits corresponding to the region of interest are connected. 

If the required connectivity within the region $\varOmega$ is not given for the graph state $\ket{\mathcal{G}}$, one can use this state to generate other locally equivalent graph states by local complementation (LC) operations: the entirety of the locally equivalent graph states $ \ket{\mathcal{G}^{(1)}},...,\ket{\mathcal{G}^{(k)}},... $ reached in this way is a finite set and corresponds to the orbit of the graph associated to the initial graph state. 

Local complementation on a site $\eta$ in a graph (represented by an adjacency matrix $\varGamma$) acts on all possible links between the sites forming the neighborhood $ \mathcal{N}_{\eta} $ of site $ \eta $ and substitutes them by their complementary, i.e~removes the existing links and adds the non-existing ones, 
\begin{equation}
\varGamma_{\mu\nu}'=\left\lbrace \begin{split} \varGamma_{\mu\nu}\;\;\;\;\;\;\;\;&\mathrm{if}\;\;\;\;\mu\;\;\mathrm{and\,/\,or}\,\;\;\nu\;\notin \mathcal{N}_{\eta}\\
\varGamma_{\mu\nu}+1  \;\;&\mathrm{if}\;\;\;\;\mu\;\;\;\;\;\mathrm{and}\,\;\;\;\;\;\nu\;\in \mathcal{N}_{\eta}
\end{split}\right.	, 
\end{equation}
where the sum is modulo 2. The unitary realizing the graph state transformation corresponding to the LC on the graph is given by a product of single-qubit unitary operations \cite{Hein2006}, 
\begin{equation}
U_{LC\eta}=\mathrm{exp}\left(-i\frac{\pi}{4}X_{\eta}\right)\prod_{\mu\in\mathcal{N}_{\eta}}\mathrm{exp}\left(i\frac{\pi}{4}Z_{\mu}\right)	.
\end{equation}
See Figures \ref{Graphs} (\textbf{a}), (\textbf{d}), (\textbf{e}) for an example of a unitary performing a LC on a graph state.

Then, for each connected subsystem in the underlying graph of each graph state obtained in this manner one can take the generator subset (Fig.~\ref{Graphs}(g)) 
\begin{equation}
	W_{\varOmega}^{\mathcal{G}^{(k)}}=\left\lbrace g_{\mu}^{(k)}:\; g_{\mu}^{(k)}\ket{\mathcal{G}^{(k)}}=\ket{\mathcal{G}^{(k)}},\;\;\mu\in\varOmega\right\rbrace
\end{equation}
and transform these graph state generators back into a subset of generators of the original stabilizer state of interest, 
\begin{equation}
W_{\varOmega}^{(k)}=U_{LE}^{-1}\left[ U_{LC}^{(k)}\right] ^{-1}W_{\varOmega}^{\mathcal{G}^{(k)}}U_{LC}^{(k)}U_{LE}, 
\end{equation}
by the inverse chain of local unitary operations corresponding to the sequence of LC operations. Due to the connectivity of the region $\varOmega$ within the graph they originate from, they satisfy Proposition \ref{p} by construction and thereby provide a valid local witness operator for state $\ket{\mathcal{S}}$ within $\varOmega$.

This approach, involving exploration of the orbit of a graph by LC operations, allows one to find local witnesses for any possible subsystem $\varOmega$ of qubits belonging to the stabilizer state. Using additional symmetries of the stabilizer state allows one to find even more witnesses: Such symmetries are local unitary operations $ U_{L\mathcal{S}}^{(t)} $, labeled by an index $ t $, which act trivially on the stabilizer state $ U_{L\mathcal{S}}^{(t)}\ket{\mathcal{S}}=\ket{\mathcal{S}} $ but are themselves not stabilizer operators. These unitaries can be found by means of solving a binary linear system in the binary picture. For instance, for the seven-qubit color code defined in Fig.~\ref{Stabilizer_state}, one of the symmetries is \begin{equation}
U_{L\mathcal{S}}=\prod_{\nu=1}^{7}\mathrm{exp}\left(i\frac{\pi}{4}X_{\nu}\right)	,
\end{equation}
which transforms the generator set $ S $ of Fig.~\ref{Stabilizer_state} into the recombined generator set $ S^{(R)} $ of Eq.~(\ref{alt_basis}). Whereas a symmetry leaves the stabilizer $ \mathcal{S} $ invariant, it may transform a generator subset $ W_{\varOmega}  $ into another one that generally will not lead to the same local witness. For instance, the generator subset $ W_{\{1,2,3,4\}} $ in Eq.~(\ref{e1}) is transformed into a generator subset
\begin{equation}
\begin{split}
	W_{\{1,2,3,4\}}'&=U_{L\mathcal{S}}W_{\{1,2,3,4\}}U_{L\mathcal{S}}^{\dagger}\\
	&=\left\lbrace s_{R}^{Z}\;,\;s^{X}_{R}s_{R}^{Z}\;,\;s^{X}_{B}s_{B}^{Z}\;,\;s^{X}_{G}s_{G}^{Z}\right\rbrace
\end{split}
\end{equation}
that constructs a different local witness because it spans a different set of stabilizers than $  W_{\{1,2,3,4\}} $:
\begin{equation}
	\left[W_{\{1,2,3,4\}}'\right] \neq\left[ W_{\{1,2,3,4\}}\right]	.
\end{equation}
Therefore, for every generator subset $ W_{\varOmega}^{(k)} $ found with this method, the generator subset $ W_{\varOmega}^{(k,t)}=\left[ U_{L\mathcal{S}}^{(t)}\right]^{-1}W_{\varOmega}^{(k)}U_{L\mathcal{S}}^{(t)} $ also constructs a local witness that may be different from the one constructed using $ W_{\varOmega}^{(k)} $ for the same subsystem. 

Interestingly, there can exist even more local witnesses that are not generated using the graph-based approach, but can be found using an alternative method outlined in the following section. 

\subsection{Direct method}\label{Dm}
\begin{figure}[t]
	\centering
	\includegraphics[width=1\columnwidth]{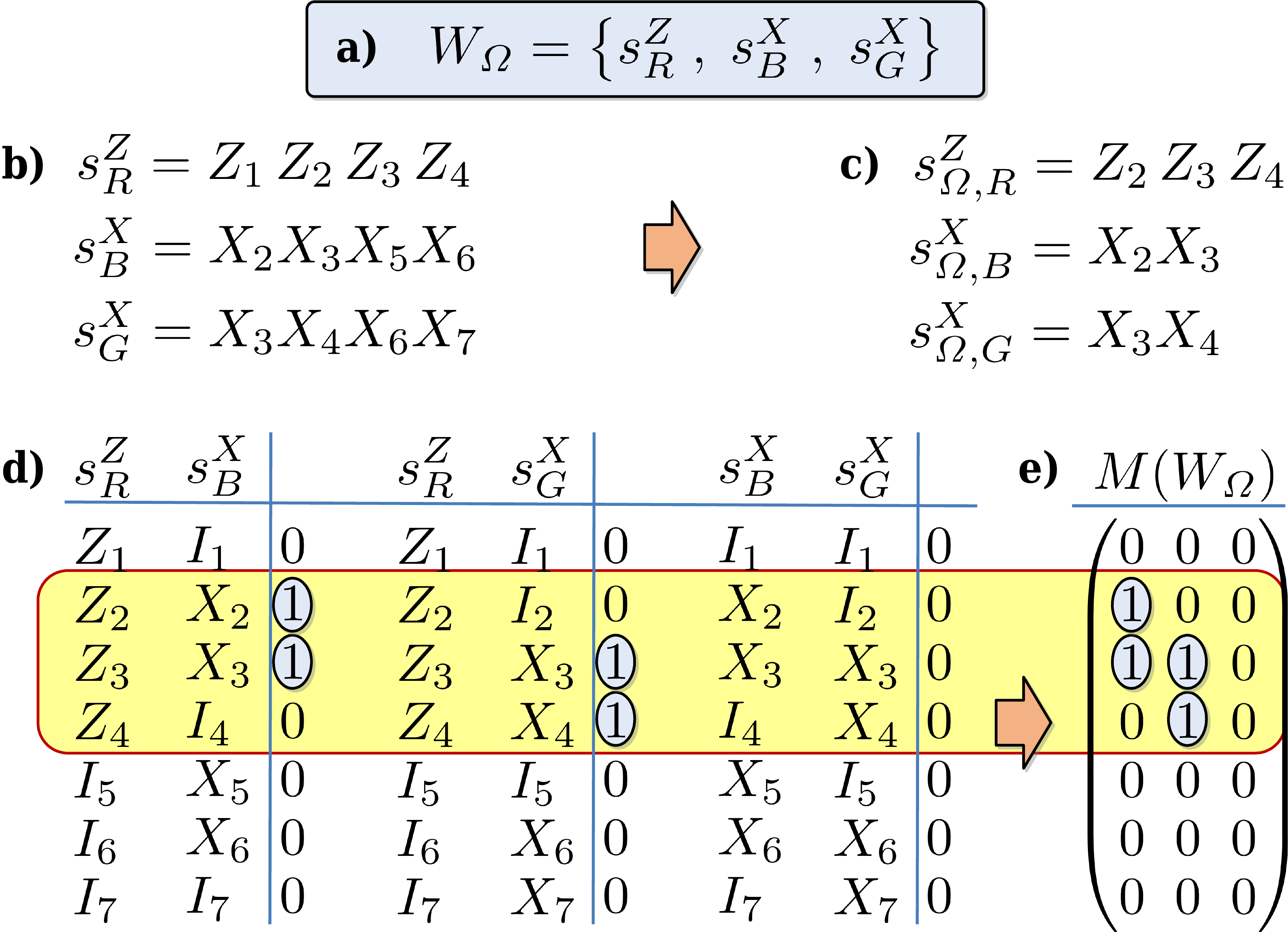}
	\caption{\textbf{Illustration of the direct method to obtain local witnesses.} The method allows one to decide if a given \textbf{(a)} generator subset $ W_{\varOmega}\subset\mathcal{S} $ composed by the stabilizers in \textbf{(b)} constructs a local witness for a subsystem $ \varOmega $. For the example shown, it contains $ n=3 $ independent and commuting stabilizers so it satisfies condition $ (i) $ in Theorem \ref{t}. \textbf{(c)} The reduced stabilizers in the subsystem $ \varOmega $ are independent and commute, thus they satisfy condition $ (ii) $. \textbf{(d)} For each pair of stabilizers $ s_{i},s_{j}\in W_{\varOmega} $ and for each qubit $ \mu $ write a $ 0 $ if the Pauli operators $ s_{i\mu},s_{j\mu}\in\{I,X,Y,Z\} $ commute $ \left[s_{i\mu},s_{j\mu}\right]=0 $, and $ 1 $ if they anti-commute $ \{s_{i\mu},s_{j\mu}\}=0 $. The subsystem $ \varOmega $ can be read from this local commutation structure: it contains the qubits $ \nu $ where at least one pair of stabilizers anti-commutes (qubits in the yellow area). \textbf{(e)} The pseudo-incidence matrix $ M\left(W_{\varOmega}\right) $ contains the information about the local commutation structure. All rows corresponding to qubits outside $ \varOmega $ are zero, thus condition $ (iii) $ holds. The rank modulo 2 of $ M\left(W_{\varOmega}\right) $ is $ n-1=2 $ here, thus condition $ (iv) $ holds. Therefore, the generator subset $ W_{\varOmega} $ constructs a valid local witness for the subsystem $ \varOmega=\{2,3,4\} $, since it satisfies $ (i),(ii),(iii),(iv) $ in Theorem \ref{t}.}
	\label{Direct_method}
\end{figure}

Here, we introduce an alternative method which provides sufficient and necessary criteria to decide if a given generator subset $ W_{\varOmega}\subset\mathcal{S} $ of stabilizers constructs a local witness for a subsystem $ \varOmega $, i.e.~that whether or not it satisfies Proposition \ref{p}. This method is illustrated in Fig.~\ref{Direct_method} with an example. Importantly, the method does not require obtaining the local unitary $ V_{LE} $, nor the recombination $ R_{\varOmega} $ of Proposition \ref{p}. Furthermore, the direct method is based on a set of intuitive criteria, which for simple cases can be checked by inspection.
\\

Given a generator subset $ W_{\varOmega}\subset\mathcal{S} $ of a stabilizer, we denote with $ s_{i\mu}  $ the single-qubit Pauli operator of the stabilizer operator $ s_{i}\in W_{\varOmega} $ on the qubit $ \mu $, 
\begin{equation}\label{paulies}
s_{i}=\bigotimes_{\mu=1}^{N}s_{i\mu}\;\;\;\;s_{i\mu}\in\left\lbrace I,X,Y,Z\right\rbrace.
\end{equation}
We will need two useful definitions: 
\begin{definition}\label{reduced_pauli_operators}
 	The $ n $ \textbf{reduced Pauli operators} $ s_{\varOmega,i} $ of a generator subset $ W_{\varOmega}\subset\mathcal{S} $ that contains $ n $ stabilizers are (Fig.~\ref{Direct_method}(b)):
 	\begin{equation}\label{reduced_stabilizers}
 		s_{\varOmega,i}=\bigotimes_{\mu\in\varOmega}s_{i\mu}\;\;\;\;\forall\;\;s_{i}\in W_{\varOmega}.
 	\end{equation}
\end{definition}
The reduced Pauli operators of the generator subset $ W_{\varOmega}^{\mathcal{G}} $ of a graph state are the reduced graph state generators $ g_{\varOmega,\mu} $ (Fig.~\ref{Graphs}(j)) defined in Eq.~(\ref{reduced_graph_generators}).

\begin{definition}\label{pseudo-incidence_matrix}
	The \textbf{\textit{pseudo-incidence matrix}} $ M(W_{\varOmega})  $ of a generator subset $ W_{\varOmega} $ of $ n $ stabilizers $ s_{i} $, is the $ N\times {n\choose 2} $ binary matrix created from the commutation or anti-commutation property of each pair of stabilizers $ s_{i},s_{j}\in W_{\varOmega} $ on each qubit (Fig.~\ref{Direct_method}(c)) as follows:
	\begin{equation}\label{ag}
	M\left( W_{\varOmega}\right)_{\mu l}=\left\lbrace 
	\begin{split}
	1 \;\;&\mathrm{if} \;\;\left\lbrace s_{i\mu},s_{j\mu}\right\rbrace=0
	\\
	0 \;\;&\mathrm{if} \;\;\;\,\left[s_{i\mu},s_{j\mu}\right]=0
	\end{split}
	\right. 	
	\end{equation}
Here, the index $ l=\left\lbrace   i,j\right\rbrace  $ labels all pairs of stabilizers $ s_{i},s_{j}\in W_{\varOmega} $ from $ 1 $ to the number of pairs $ {n\choose 2} $, and the index $ \mu $ labels qubits from $ 1 $ to $ N $.
\end{definition}
Note that the pseudo-incidence matrix is preserved under local unitary operations, 
\begin{equation}
M\left( V_{LE}W_{\varOmega}V_{LE}^{\dagger}\right) =M\left( W_{\varOmega}\right), 
\end{equation}
because commutation and anti-commutation relations of single-qubit Pauli operators $ s_{i\mu} $ on each qubit are preserved under these operations.

The following theorem then provides the necessary and sufficient criteria to decide whether a generator subset constructs a local witness for a subsystem or not:
\begin{theorem}\label{t}
	The generator subset $ W_{\varOmega}\subset\mathcal{S} $ satisfies Proposition \ref{p} i.e.~constructs a local witness for the subsystem $ \varOmega $ if and only if it satisfies the following properties:
	\begin{enumerate}[label=(\roman*)]
		\item $ W_{\varOmega} $ contains $ n $ independent and commuting stabilizers.
		\item The $ n $ reduced Pauli operators $ s_{\varOmega,i} $ in $ \varOmega $ are independent and commute.
		\item The rows of the pseudo-incidence matrix $ M\left( W_{\varOmega}\right)  $ corresponding to the qubits outside $ \varOmega $ are zero, i.e.~$M\left( W_{\varOmega}\right)_{\eta l}=0 \,\,\forall l$ and $\forall \eta\notin\varOmega$.
% 		\begin{equation}
%		M\left( W_{\varOmega}\right)_{\eta l}=0\;\;\;\;\forall\;\;l\;\;\mathrm{and}\;\;\forall\;\;\eta\notin\varOmega	.
%		\end{equation}
		\item The modulo 2 rank of the pseudo-incidence matrix $M\left( W_{\varOmega}\right) = n-1$.
%		\begin{equation}
%		\mathrm{rank}\left( M\left( W_{\varOmega}\right) \right) =n-1	.
%		\end{equation}
	\end{enumerate}
\end{theorem}

To prove this we first show that $ (i),(ii),(iii) $ are sufficient and necessary for $ (A) $ in Proposition \ref{p}. Suppose that the generator subset $ W_{\varOmega} $ satisfies $ (A) $. Then, the generator subset $ W_{\varOmega}^{\mathcal{G}} $ exists and contains $ n $ independent and commuting graph state generators. This is preserved under any local unitary and recombinations, so $ W_{\varOmega} $ contains $ n $ independent and commuting stabilizers. Then $ (A) $ implies $ (i) $. The same applies to the reduced Pauli operators: the $ n $ reduced graph state generators of $ W_{\varOmega}^{\mathcal{G}} $ are independent and commute, thus the same holds for the reduced Pauli operators of $ W_{\varOmega} $. Then $ (A) $ implies $ (ii) $. The single-qubit Pauli operators of $ W_{\varOmega}^{\mathcal{G}} $ on qubits outside $ \varOmega $ are of two types only, namely $ I $ and $ Z $, thus they commute. Again, this is preserved by local unitary operations and recombinations, therefore the single-qubit Pauli operators of $ W_{\varOmega} $ on these qubits are of two types, $ I_{\eta} $ and $ \sigma_{\eta} $, only, which is the same for each qubit $ \eta\notin \varOmega $. They commute on each qubit, thus the rows $ \eta\notin \varOmega $ of $ M\left( W_{\varOmega}\right)  $ are zero. Then $ (A) $ implies $ (iii) $.

To show the converse, now suppose that the generator subset $ W_{\varOmega} $ satisfies $ (i),(ii),(iii) $. If $ (i) $, the generator subset $ V_{LE}W_{\varOmega}V_{LE}^{\dagger} $ contains $ n $ independent and commuting stabilizers. If $ (ii) $ holds, the $ n $ reduced Pauli operators $ V_{LE}s_{\varOmega,i}V_{LE}^{\dagger} $ form a complete generator set of the reduced graph state $ \ket{\mathcal{G}} $. It was shown in \cite{Van2004} that there is always a local unitary $ V_{LE,\varOmega} $ and a recombination $ R_{\varOmega} $ that transforms them into graph state generators, 
\begin{equation}
	V_{LE,\varOmega}\prod_{j=1}^{n}s_{\varOmega,i}^{\left[ R_{\varOmega}\right] _{ij}}V_{LE,\varOmega}^{\dagger}=g_{\varOmega,\mu}\;\;\;\;\forall\;\;s_{\varOmega,i}, 
\end{equation}
where $ g_{\varOmega,\mu} $ are the reduced graph state generators of the reduced graph state $ \ket{\mathcal{G}_{\varOmega}} $ and the mapping $ i\mapsto\mu $ is one-to-one.

As an example, in Eq.~(\ref{mapping}), the reduced Pauli operators of the generator subset of Eq.~(\ref{e1}) are transformed into reduced graph state generators
\begin{equation}
\begin{split}
	s_{\varOmega,1}=Y_{1}Y_{2}Y_{3}Y_{4}\;\;&\mapsto\;\;g_{\varOmega,3}=Z_{1}Z_{2}X_{3}Z_{4}\\
	s_{\varOmega,2}=Z_{1}Z_{2}Z_{3}Z_{4}\;\;&\mapsto\;\;g_{\varOmega,1}=X_{1}Z_{3}\\
	s_{\varOmega,3}=Z_{2}Z_{3}\;\;&\mapsto\;\;g_{\varOmega,2}=X_{2}Z_{3}\\
	s_{\varOmega,4}=Z_{3}Z_{4}\;\;&\mapsto\;\;g_{\varOmega,4}=Z_{3}X_{4}
\end{split}
\end{equation}
with the local unitary $ V_{LE,\varOmega}=H_{1}H_{2}H_{4} $ and the recombination
\begin{equation}
R_{\varOmega}=\begin{pmatrix}
1&0&0&0\\
1&1&0&0\\
0&1&1&0\\
0&1&0&1\\
\end{pmatrix}.
\end{equation}

Finally, if $ (iii) $ holds, the single-qubit Pauli operators $ s_{i\eta} $ on each qubit $ \eta\notin\varOmega $ can only be of two types: $ I_{\eta} $ and $ \sigma_{\eta} $. Then, there exists always a local unitary $ V_{LE,\bar{\varOmega}} $ that transforms them into $ I_{\eta} $ and $ Z_{\eta} $. For instance, for the generator subset in Fig.~\ref{Direct_method}(a) this unitary is given by $ V_{LE,\bar{\varOmega}}=H_{5}H_{6}H_{7} $. 

The conclusion is that if the generator subset $ W_{\varOmega} $ satisfies $ (i),(ii),(iii) $, the local unitary $ V_{LE}=V_{LE,\varOmega}\otimes V_{LE,\bar{\varOmega}} $ and the recombination $ R_{\varOmega} $ exist, and thus, $ (i),(ii),(iii) $ implies $ (A) $ in Proposition \ref{p}.
\\
To complete the proof we now show that if the generator subset $ W_{\varOmega} $ that satisfies $ (A) $, or equivalently, $ (i),(ii),(iii) $, then $ (B) $ and $ (iv) $ are equivalent.

Consider first the pseudo-incidence matrix $ M(G) $ of a complete generator set $ G $ of $ N $ generators of a graph state $ \ket{\mathcal{G}} $. One can check that it coincides with the incidence matrix $ M_{E} $ of the underlying graph. First, the column $ l=\{\mu,\nu\} $ that compares the graph state generators $ g_{\mu},g_{\nu} $ is either zero or contains two ones. It is zero only if the pair of sites $ l $ is not in the graph ($ l\notin E $). But if the two sites are linked $ l\in E $, the graph state generators anti-commute precisely on the qubits corresponding to those sites. Then, $ \mathrm{rank}\left(M(G)\right) = N - 1 $ if and only if $ \ket{\mathcal{G}} $ is non-separable. The same happens with a generator set $ G_{\varOmega} $ of $ n $ reduced graph state generators $ g_{\varOmega,\mu} $: $ \mathrm{rank}\left(M(G_{\varOmega})\right) =n-1 $ if and only if $ \ket{\mathcal{G}_{\varOmega}} $ is non-separable.

Now, given that the non-zero rows of the pseudo-incidence matrix $ M\left(W^{\mathcal{G}}_{\varOmega}\right) $ coincide with the rows of $ M(G_{\varOmega})$, their ranks also coincide. Additionally, the local unitary operations preserve the pseudo-incidence matrix, thus we can conclude that $ \mathrm{rank}\left(M(W_{\varOmega}^{(R_{\varOmega})})\right) =n-1 $ if and only if $ \ket{\mathcal{G}_{\varOmega}} $ is non-separable.

In Appendix \ref{recombinations} we show that an invertible recombination $ R_{\varOmega} $ of the stabilizers in $ W_{\varOmega} $ induces an invertible recombination $ \tilde{R} $ of the columns of the pseudo-incidence matrix, 
\begin{equation}\label{pseudo_recombined}
	M\left( W_{\varOmega}^{(R_{\varOmega})}\right) =M\left( W_{\varOmega}\right)\tilde{R}, 
\end{equation}
where the matrix product between the binary matrices $ M\left( W_{\varOmega}\right) $ and the binary non-singular matrix $ \tilde{R} $ is performed modulo 2. Thus, the rank of the pseudo-incidence matrix is preserved under recombinations, confirming that properties $ (B) $ and $ (iv) $ are equivalent, thus, completing the proof of Theorem \ref{t}.

As mentioned earlier, the direct method allows one to find more local witnesses than the graph-based method because the condition on the local unitary $ V_{LE} $ is less stringent than the condition on the local unitary $ U_{LE} $ used in the graph-based method. On the one hand, the local unitary $ V_{LE} $ must transform $ W_{\varOmega} $ into some $ W^{\mathcal{G}}_{\varOmega} $ up to a recombination, but it is not forced to transform the stabilizer state $ \ket{\mathcal{S}} $ into a graph state $ \ket{\mathcal{G}} $. On the other hand, $ U_{LE} $ must convert $ \ket{\mathcal{S}} $ into $ \ket{\mathcal{G}} $, and this always guarantees that it converts $ W_{\varOmega} $ into some $ W^{\mathcal{G}}_{\varOmega} $ up to a recombination. For example, the witness constructed with the generator subset used as example in Fig.~\ref{Direct_method}(a) is found with the direct method but it cannot be found with the graph-based method (see Appendix \ref{non_graph-based_wit}). In this example, the unitary $ V_{LE}=H_{3}H_{5}H_{6}H_{7} $ satisfies property $ (A) $ but one can check that $ V_{LE}\ket{\mathcal{S}} $ is not a graph state, and furthermore, there is no local unitary that satisfies $ (A) $ and also converts the seven-qubit color code into some graph state.

\section{Local witnesses on an experimental state: the seven-qubit color code}\label{S5}
We now examine the performance of the developed local witness operators on a concrete example of a recently implemented experimental seven-qubit color code state \cite{Nigg2014}. This multi-qubit state provides a rich playground to explore interesting features of the local witnesses introduced and their practical suitability for entanglement characterization To this end, we construct all local witnesses that can be constructed with both methods for all the 119 possible subsystems of the seven-qubit color code evaluated them based on the experimental data of Ref.~\cite{Nigg2014}. These 119 subsystems are simply the $ 2^7=128 $ possible subsets of seven qubits excluding the empty set, the subset containing the seven qubits, and the seven subsets containing only one qubit.

\begin{table}[t]
	\centering
	\renewcommand{\arraystretch}{2}
	
	\begin{tabular}{ C{3em}  || C{9em} | C{4em} C{4em} C{4em}}
		$ n $ & Subsystem $ \varOmega $ & Graph-based & Direct method & Two-meas.\\
		\hline \hline
		2  &  \cellcolor[HTML]{F9C7A0} All  & \cellcolor[HTML]{F9C7A0} 54   & \cellcolor[HTML]{F9C7A0} 72  & \cellcolor[HTML]{F9C7A0} 4 \\
		\hline
		\multirow{2}{*}{3} &  \cellcolor[HTML]{F6DCC9}	String like & \cellcolor[HTML]{F6DCC9}  32     & \cellcolor[HTML]{F6DCC9} 40   & \cellcolor[HTML]{F6DCC9} 4   \\& 
		\cellcolor[HTML]{F9C7A0} Non string like   & \cellcolor[HTML]{F9C7A0}  34 & \cellcolor[HTML]{F9C7A0} 44  & \cellcolor[HTML]{F9C7A0} 5  \\
		\hline
		\multirow{2}{*}{4} & 	\cellcolor[HTML]{F6DCC9} Plaquette like &  \cellcolor[HTML]{F6DCC9} 17  \cellcolor[HTML]{F6DCC9}   & \cellcolor[HTML]{F6DCC9} 30   & \cellcolor[HTML]{F6DCC9} 9   \\& 
		\cellcolor[HTML]{F9C7A0} Non plaquette like   &  \cellcolor[HTML]{F9C7A0} 18 & \cellcolor[HTML]{F9C7A0} 18  & \cellcolor[HTML]{F9C7A0} 3  \\
		\hline
		5 & \cellcolor[HTML]{F6DCC9} All & \cellcolor[HTML]{F6DCC9} 8   & \cellcolor[HTML]{F6DCC9} 8 & \cellcolor[HTML]{F6DCC9} 3 \\
		\hline
		6 & \cellcolor[HTML]{F9C7A0} All & \cellcolor[HTML]{F9C7A0} 3 \cellcolor[HTML]{F9C7A0}  & \cellcolor[HTML]{F9C7A0} 3 & \cellcolor[HTML]{F9C7A0} 2 \\
		\hline \hline
		Total & 119 & 3122 & 3927 & 476
	\end{tabular}
	\caption{\textbf{Local witnesses for the seven-qubit color code}. Columns from left to right: number of qubits in the subsystem, classification of subsystems, number of standard local witnesses constructed with the graph-based method, number of standard local witnesses constructed with the direct method, and number of two-measurements local witnesses.}
	\label{t2}
\end{table}

\begin{figure*}[t]
	\centering
%	\hspace*{-1.5cm}
	\includegraphics[width=2\columnwidth]{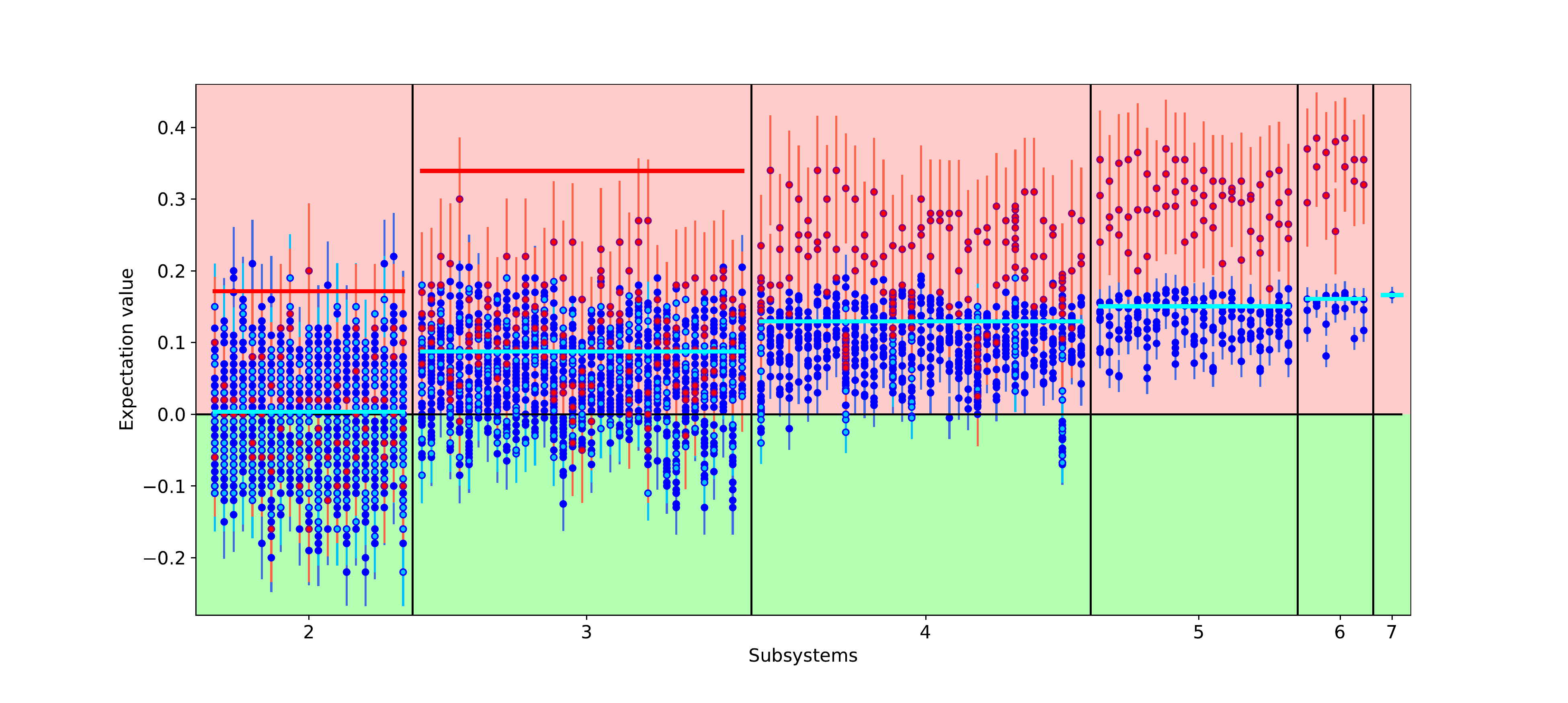}
	\caption{\textbf{Witnesses expectation values}. Each point corresponds to the expectation value of a witness evaluated on the experimental implementation of the seven-qubit color code \cite{Nigg2014}, so each point in the green area is an entanglement detection. The subsystems $ \varOmega $ are ordered in 119 columns from left to right according to the following convention $ \{1,2\} $, $ \{1,3\} $, ..., $ \{1,7\} $, $ \{2,3\} $, ..., $ \{2,7\} $, ..., $ \{6,7\} $ for the two-qubit subsystems, then followed by $ \{1,2,3\} $, ..., $ \{1,2,7\} $, $ \{1,3,4\} $, ..., $ \{1,3,7\} $, ..., $ \{2,3,4\} $, ..., $ \{5,6,7\} $ as the three-qubit subsystems, and so forth, up to the last six-qubit subsystem $ \{2,3,4,5,6,7\} $. The point in the very last column corresponds to the standard genuine witnesses (the value of the two-measurements genuine witness is 0.52(5), which lies outside the scope of the plot and is not shown). Light blue points correspond to the local witnesses obtained only with the direct method, and the dark blue points to witnesses obtained with both the direct method and the graph-based method. Red points correspond to two-measurement local witnesses. The horizontal blue and red lines are the positions where the standard local witnesses and the two-measurements local witnesses, respectively, would lie if the state was accurately described by the simplified Werner state model, adjusted to reproduce the experimental fidelity of 0.33(1). Missing red lines for $n >3$ subsets lie outside the shown range of the figure and not shown.}
	\label{Big_scatter_plot}
\end{figure*}

\subsection{Constructed local witnesses}
We generate the entanglement witnesses systematically using a computer program and find that the graph-based method provides 3122 standard local witnesses whereas the direct method provides a larger number of 3927 witnesses. The difference comes from the fact that not all local witnesses for a stabilizer state can be constructed from the orbit of local unitary equivalent graph states. Note that the 3927 standard local witnesses constitute all witnesses of this structure that exist for the seven-qubit color code state. This is because the direct method provides the sufficient and necessary conditions for them to exist and we go through all possible generator subsets $W_{\varOmega}\subset\mathcal{S}$ that can be formed out of the $2^7 -1 =127$ non-trivial stabilizers, checking for each subset whether it satisfies the criteria or not. From the generator subsets which give rise to these 3927 local witnesses we also construct all 476 possible two-measurements local witnesses. The results are summarized in Table \ref{t2}. We find, as expected, that the three groups of local witnesses contain at least one local witness for each of the possible 119 subsystems. 

The distribution of local witness operators associated to the 119 possible subsystems results from the structure of the ideal seven-qubit color code state. It is useful to distinguish between two types of three-qubit subsystems and two types of 4-qubit subsystems. The ``string-like subsystems" are three-qubit subsystems in which the multiplication of the logical operator $ s_{L}^{X} $ with any combination of $ X $-type plaquette operator has support on one of the following three-qubit sets, $ \left\lbrace 1,2,5\right\rbrace  $, $ \left\lbrace 1,4,7\right\rbrace  $, $ \left\lbrace 5,6,7\right\rbrace  $, $ \left\lbrace 1,3,6\right\rbrace  $, $ \left\lbrace 3,4,5\right\rbrace  $, $ \left\lbrace 2,3,7\right\rbrace  $, and $ \left\lbrace 2,4,6\right\rbrace  $, while the ``non-string-like subsystems" are the rest of three-qubit subsystems. Similarly, the ``plaquette-like subsystems" are four-qubit subsystems in which the multiplication of the $ Z $-type stabilizers has support, $ \left\lbrace 1,2,3,4\right\rbrace  $, $ \left\lbrace 2,3,5,6\right\rbrace  $, $ \left\lbrace 3,4,6,7\right\rbrace  $, $ \left\lbrace 1,2,6,7\right\rbrace  $, $ \left\lbrace 1,4,5,6\right\rbrace  $, $ \left\lbrace 2,4,5,7\right\rbrace  $, and $ \left\lbrace 1,3,5,7\right\rbrace  $, whereas ``non-plaquette operator like subsystems" refer to the rest of four-qubit subsystems.

\subsection{Evaluation of experimental data}
In a recent experiment, the seven-qubit color code was encoded using a linear ion-trap quantum processor formed of a string of seven $ ^{40}\mathrm{Ca}^{+} $ ions \cite{Nigg2014}. The experimental state was realized unitarily by successive creation of entanglement on each plaquette using four-qubit entangling operations combined with sequences of single-qubit operations to spectroscopically decouple and activate physical qubits - we refer the interested reader for details to Ref.~\cite{Nigg2014}.
\begin{table*}[]
	\centering
	\renewcommand{\arraystretch}{1.5}
	\begin{tabular}{C{5em} || C{1.7em} | C{1em} C{1em} C{1em} C{1em} C{1em} C{1em} C{1em} C{1em} C{1em} C{1em} C{1em} C{1.7em} | C{1em} C{1em} C{1em} C{1em} C{1em} C{1em} || C{1em} C{1em} C{1em} C{1em} C{1em} C{1em} C{1.7em} | C{1em} C{1em} C{1em} C{1em} C{1em} C{1em}}
	& \multicolumn{19}{c||}{\textbf{Standard local witnesses detections}}		& \multicolumn{13}{c}{\textbf{Two-meas. local witnesses detections}} \\ \hline \hline
	Confi. (\%)   & \cellcolor[HTML]{9AFF99} $ > $97	& \cellcolor[HTML]{CCFFCC}73	& \cellcolor[HTML]{9AFF99}84 & \cellcolor[HTML]{CCFFCC}55 & \cellcolor[HTML]{9AFF99}87 & \cellcolor[HTML]{CCFFCC}84 & \cellcolor[HTML]{9AFF99}77 & \cellcolor[HTML]{CCFFCC}81 & \cellcolor[HTML]{9AFF99}60 & \cellcolor[HTML]{CCFFCC}87 & \cellcolor[HTML]{9AFF99}73 & \cellcolor[HTML]{CCFFCC}69 & \cellcolor[HTML]{9AFF99}$ > $55 & \cellcolor[HTML]{CCFFCC}92 & \cellcolor[HTML]{9AFF99}75 & \cellcolor[HTML]{CCFFCC}80 & \cellcolor[HTML]{9AFF99}57 & \cellcolor[HTML]{CCFFCC}57 & \cellcolor[HTML]{9AFF99}99 & \cellcolor[HTML]{CCFFCC}77 & \cellcolor[HTML]{9AFF99}59 & \cellcolor[HTML]{CCFFCC}69 & \cellcolor[HTML]{9AFF99}77 & \cellcolor[HTML]{CCFFCC}69 & \cellcolor[HTML]{9AFF99}69 & \cellcolor[HTML]{CCFFCC}$ > $59 & \cellcolor[HTML]{9AFF99}55 & \cellcolor[HTML]{CCFFCC}71 & \cellcolor[HTML]{9AFF99}75 & \cellcolor[HTML]{CCFFCC}56 & \cellcolor[HTML]{9AFF99}75 & \cellcolor[HTML]{CCFFCC}66                \\ \hline
	
	 & \cellcolor[HTML]{9AFF99} & \cellcolor[HTML]{CCFFCC}1 & \cellcolor[HTML]{9AFF99}1 & \cellcolor[HTML]{CCFFCC}1 & \cellcolor[HTML]{9AFF99}2 & \cellcolor[HTML]{CCFFCC}2 & \cellcolor[HTML]{9AFF99}2 & \cellcolor[HTML]{CCFFCC}2 & \cellcolor[HTML]{9AFF99}2 & \cellcolor[HTML]{CCFFCC}3 & \cellcolor[HTML]{9AFF99}3 & \cellcolor[HTML]{CCFFCC}4 & \cellcolor[HTML]{9AFF99} & \cellcolor[HTML]{CCFFCC}1 & \cellcolor[HTML]{9AFF99}1 & \cellcolor[HTML]{CCFFCC}1 & \cellcolor[HTML]{9AFF99}1 & \cellcolor[HTML]{CCFFCC}2 & \cellcolor[HTML]{9AFF99}3 & \cellcolor[HTML]{CCFFCC}1 & \cellcolor[HTML]{9AFF99}2 & \cellcolor[HTML]{CCFFCC}3 & \cellcolor[HTML]{9AFF99}4 & \cellcolor[HTML]{CCFFCC}4 & \cellcolor[HTML]{9AFF99}5 & \cellcolor[HTML]{CCFFCC} & \cellcolor[HTML]{9AFF99}1 & \cellcolor[HTML]{CCFFCC}2 & \cellcolor[HTML]{9AFF99}2 & \cellcolor[HTML]{CCFFCC}2 & \cellcolor[HTML]{9AFF99}2 & \cellcolor[HTML]{CCFFCC}3 \\ 
	 
	Subsystem & \multirow{-2}{*}{\(\displaystyle \subset \)}\cellcolor[HTML]{9AFF99} & \cellcolor[HTML]{CCFFCC}3 & \cellcolor[HTML]{9AFF99}4 & \cellcolor[HTML]{CCFFCC}5 & \cellcolor[HTML]{9AFF99}3 & \cellcolor[HTML]{CCFFCC}4 & \cellcolor[HTML]{9AFF99}4 & \cellcolor[HTML]{CCFFCC}5 & \cellcolor[HTML]{9AFF99}5 & \cellcolor[HTML]{CCFFCC}5 & \cellcolor[HTML]{9AFF99}5 & \cellcolor[HTML]{CCFFCC}5 & \cellcolor[HTML]{9AFF99}\multirow{-2}{*}{\(\displaystyle \subset \)} & \cellcolor[HTML]{CCFFCC}2 & \cellcolor[HTML]{9AFF99}2 & \cellcolor[HTML]{CCFFCC}2 & \cellcolor[HTML]{9AFF99}4 & \cellcolor[HTML]{CCFFCC}3 & \cellcolor[HTML]{9AFF99}4 & \cellcolor[HTML]{CCFFCC}6 & \cellcolor[HTML]{9AFF99}4 & \cellcolor[HTML]{CCFFCC}4 & \cellcolor[HTML]{9AFF99}5 & \cellcolor[HTML]{CCFFCC}6 & \cellcolor[HTML]{9AFF99}7 & \cellcolor[HTML]{CCFFCC}\multirow{-2}{*}{\(\displaystyle \subset \)} & \cellcolor[HTML]{9AFF99}2 & \cellcolor[HTML]{CCFFCC}3 & \cellcolor[HTML]{9AFF99}3 & \cellcolor[HTML]{CCFFCC}3 & \cellcolor[HTML]{9AFF99}6 & \cellcolor[HTML]{CCFFCC}5 \\
	
	\(\displaystyle \varOmega \) & \cellcolor[HTML]{9AFF99} & \cellcolor[HTML]{CCFFCC}6 & \cellcolor[HTML]{9AFF99}7 & \cellcolor[HTML]{CCFFCC}7 & \cellcolor[HTML]{9AFF99}6 & \cellcolor[HTML]{CCFFCC}6 & \cellcolor[HTML]{9AFF99}7 &  \cellcolor[HTML]{CCFFCC}6 & \cellcolor[HTML]{9AFF99}7 & \cellcolor[HTML]{CCFFCC}6 & \cellcolor[HTML]{9AFF99}7 & \cellcolor[HTML]{CCFFCC}7 & \cellcolor[HTML]{9AFF99} & \cellcolor[HTML]{CCFFCC}3 & \cellcolor[HTML]{9AFF99}3 & \cellcolor[HTML]{CCFFCC}6 & \cellcolor[HTML]{9AFF99}5 & \cellcolor[HTML]{CCFFCC}4 & \cellcolor[HTML]{9AFF99}6 & \cellcolor[HTML]{CCFFCC} & \cellcolor[HTML]{9AFF99} & \cellcolor[HTML]{CCFFCC} & \cellcolor[HTML]{9AFF99} & \cellcolor[HTML]{CCFFCC} & \cellcolor[HTML]{9AFF99} & \cellcolor[HTML]{CCFFCC} & \cellcolor[HTML]{9AFF99}7 & \cellcolor[HTML]{CCFFCC}5 & \cellcolor[HTML]{9AFF99}6 & \cellcolor[HTML]{CCFFCC}7 & \cellcolor[HTML]{9AFF99}7 & \cellcolor[HTML]{CCFFCC}6 \\
	
	& \cellcolor[HTML]{9AFF99} & \cellcolor[HTML]{CCFFCC} & \cellcolor[HTML]{9AFF99} & \cellcolor[HTML]{CCFFCC} & \cellcolor[HTML]{9AFF99} & \cellcolor[HTML]{CCFFCC} & \cellcolor[HTML]{9AFF99} & \cellcolor[HTML]{CCFFCC} & \cellcolor[HTML]{9AFF99} & \cellcolor[HTML]{CCFFCC} & \cellcolor[HTML]{9AFF99} & \cellcolor[HTML]{CCFFCC} & \cellcolor[HTML]{9AFF99} & \cellcolor[HTML]{CCFFCC}4 & \cellcolor[HTML]{9AFF99}7 & \cellcolor[HTML]{CCFFCC}7 & \cellcolor[HTML]{9AFF99}6 & \cellcolor[HTML]{CCFFCC}5 & \cellcolor[HTML]{9AFF99}7 & \cellcolor[HTML]{CCFFCC} & \cellcolor[HTML]{9AFF99} & \cellcolor[HTML]{CCFFCC} & \cellcolor[HTML]{9AFF99} & \cellcolor[HTML]{CCFFCC} & \cellcolor[HTML]{9AFF99} & \cellcolor[HTML]{CCFFCC} & \cellcolor[HTML]{9AFF99} & \cellcolor[HTML]{CCFFCC} & \cellcolor[HTML]{9AFF99} & \cellcolor[HTML]{CCFFCC} & \cellcolor[HTML]{9AFF99} & \cellcolor[HTML]{CCFFCC} \\ \hline
	Size of $\varOmega$ & 2 & \multicolumn{12}{c|}{3} & \multicolumn{6}{c||}{4} & \multicolumn{7}{c|}{2} & \multicolumn{6}{c}{3}                                                                                                                                                                                                                                                                                                                                                                                                                                
\end{tabular}	
\caption{\textbf{Local entanglement detections}. The rows contain from top to bottom: the type of witness, the confidence (in \%) in the fact that the most negative expectation value of all the witnesses for the same subsystem $ \varOmega $ (in the same column of Fig.~\ref{Big_scatter_plot}) is below $ 0 $, the subsystem $ \varOmega $, and number $ n $ of qubits in each $ \varOmega $. The columns with the subset symbol $ (\subset) $ represents all subsystems $ \varOmega' $ inside at least one of the subsystems $ \varOmega $ presented in the columns at the right of the symbol. For instance, the first column of $ n=2 $ represents the 21 two-qubit subsystems inside the presented three and four-qubit subsystems. To obtain confidence of this column we compare the most negative witnesses for each subsystem $ \varOmega' $ and then take the confidence of the most positive one among them.}
	\label{Detections}
\end{table*}

Perfect implementation of the encoding sequence would lead to the ideal genuinely entangled state with an expectation value of all witnesses of $ -1/2 $. However, due to state preparation, gate and measurement errors, the quantum state fidelity of the experimental state with the expected state was limited to $ 0.33(1) $, illustrating the usefulness of witness operators to reveal the structure of local quantum correlations in such noisy multi-qubit state. Figure~\ref{Big_scatter_plot} shows the expectation values of the genuine and local witnesses, as calculated based on the experimental measurement data for the 127 non-trivial stabilizer operators. Error bars correspond to the variance based on the assumption of independent binomial distributions of expectation values for each stabilizer operator (see Appendix \ref{bin_distr} for details).

The first observation is that we detect entanglement in multiple subsystems, which are summarized in Table \ref{Detections}. The detections by the standard local witnesses are for all $7\cdot 6/2 = 21$ two-qubit subsystems, all 35 three-qubit subsystems except $ \{1,2,5\} $ and $ \{1,3,5\} $, and 6 four-qubit subsystems out of the possible 35. The detections of the two-measurements local witnesses are: all the 21 two-qubit subsystems except $ \{1,3\} $, $ \{1,4\} $, and $ \{4,7\} $, and 6 three-qubit subsystems out of the possible 35. Regarding the distribution of points the results of Fig.~\ref{Big_scatter_plot} confirm what Eq.~(\ref{finer}) suggests, namely that local witnesses for small subsystems tend to have a more negative expectation value, while local witnesses for bigger subsystems hardly detect entanglement. In fact, for every subsystem $ \varOmega' $ where entanglement is detected, all subsystems $ \varOmega\subset\varOmega' $ of it present entanglement as well. For instance, all two-qubit and three-qubit subsystems with qubits only in the red plaquette show entanglement because 4-qubit entanglement in subsystem $ \{1,2,3,4\} $ has been detected.

It can also be seen that as expected the two-measurements witnesses tend to have more positive values than the standard ones, in accordance with Eq.~(\ref{finnermodified}), and they also become even more positive as the number of qubits in the subsystem increases. Therefore, with two-measurements local witnesses one can detect the entanglement of small subsystems with just two measurement settings, for the state at the given noise levels, but for larger subsystems the standard local witnesses, requiring more measurement settings provide a more reliable choice to detect entanglement. 

A further interesting observation is that the Werner model can at best be a rough approximation of the experimental state at hand. In this model, all stabilizers have the same expectation value $ p $, and this value coincides with the fidelity between the Werner state and the stabilizer state. Thus, in order to compare the Werner state model with the experimental state we set the value of $ p $ as so it reproduces the experimental state fidelity of 0.33(1). The resulting expectation value of all local witnesses is positive when evaluated on the Werner model, so no entanglement is detected. However, when evaluated on the experimental state, different stabilizers have varying expectation values. Then, local witnesses constructed predominantly from stabilizers with an expectation value above the fidelity have an expectation value below the prediction of the Werner model. Some of these witnesses have negative expectation values significantly below zero when evaluated on the experimental state, resulting in the detection of entanglement in the respective subsets of qubits.

\section{Conclusions and Outlook}\label{S6}
In this work we have proposed and explored two methods to construct local witness operators for stabilizer states, which can be decomposed into stabilizer operators of corresponding state of interest. We have shown how these operators, detecting the entanglement among qubits in subsystems of larger multi-qubit states, allow one to resolve the entanglement structure in experimental stabilizer states. In our work, we have built on ideas proposed in \cite{Alba2010} for graph states and extended this to arbitrary subsets of qubits in graph states and general stabilizer states. Additionally, we have explored a so-called direct method which provides sufficient and necessary criteria to decide if a local witness can be constructed from a generator subset, and which, in general, provides a larger set of local witness operators than the graph-based technique. Following the spirit of Ref.~\cite{Toth2005_94}, we have also constructed modified local witnesses that require less measurement settings, at the cost of a decrease in sensitivity to detect entanglement in noisy states. Finally, we have illustrated the construction of the witnesses and studied their practical suitability for local entanglement detection by applying them to an experimental implementation of a 7-qubit quantum error correcting code \cite{Nigg2014}.

Possible future extensions of the present work could include generalizations of the methods to construct local witnesses to non-stabilizer states like  e.g.~W-states \cite{Zeilinger1992, Dur2000} and Dicke states \cite{Dicke1954}, and the exploration of the utility of such local entanglement witnesses in the context of detecting phase transitions in condensed matter systems.

\section*{Acknowledgments}
We thank D. Nigg, E. Martinez and P. Schindler for useful discussions, and the authors of Ref.~\cite{Nigg2014} for providing us with access to the measurement data. We acknowledge support by U.S. A.R.O. through Grant No. W911NF-14-1-010. M.M.~acknowledges support from the AQTION project (820495), funded by the European Union Quantum Technology Flagship. The research is also based upon work supported by the Office of the Director of National Intelligence (ODNI), Intelligence Advanced Research Projects Activity (IARPA), via the U.S. Army Research Office Grant No. W911NF-16-1-0070. The views and conclusions contained herein are those of the authors and should not be interpreted as necessarily representing the official policies or endorsements, either expressed or implied, of the ODNI, IARPA, or the U.S. Government. The U.S. Government is authorized to reproduce and distribute reprints for Governmental purposes notwithstanding any copyright annotation thereon. Any opinions, findings, and conclusions or recommendations expressed in this material are those of the author(s) and do not necessarily reflect the view of the U.S. Army Research Office.

\appendix
\clearpage

\section{Binary picture of the stabilizer formalism}\label{BIN_PIC}
Here we introduce for completeness the binary picture of the stabilizer formalism. In the binary picture, stabilizers and the Clifford group acting on the Hilbert space $ \mathcal{H}_{2}^{N} $ are represented by binary vectors and matrices, respectively \cite{Van2004,Hein2006}. We will follow the notation used in Ref.~\cite{Van2004}.

In the \textit{binary picture} of the stabilizer formalism for one qubit $ \mathcal{H}_{2}^{1} $ the single-qubit Pauli operators are $ 2\times 1 $ column vectors:
\begin{equation}
I\mapsto\begin{pmatrix}
0\\\hline 0
\end{pmatrix}\;,\;	X\mapsto\begin{pmatrix}
0\\\hline 1
\end{pmatrix}\;,\;	Y\mapsto\begin{pmatrix}
1\\\hline 1
\end{pmatrix}\;,\;	Z\mapsto\begin{pmatrix}
1\\\hline 0
\end{pmatrix}
\end{equation}
and matrix multiplication is transformed into a vector addition modulo 2:
\begin{equation}
XY=iZ\;\;\mapsto\;\;\begin{pmatrix}
0\\\hline 1
\end{pmatrix}+\begin{pmatrix}
1\\\hline 1
\end{pmatrix}=\begin{pmatrix}
1\\\hline 0
\end{pmatrix}	.
\end{equation}
We disregard the phases $ \pm1,\,\pm i $ because in all cases relevant for the present work they result in global phases with no influence.  

In the binary picture of $ N $ qubits the stabilizers $ s_{i} $ formed by single-qubit Pauli operators $ s_{i\mu} $ (see Eq.~(\ref{paulies})) are represented by $ 2N\times 1 $ column vectors:
\begin{equation}
	\mathbb{s}_{i}=\begin{pmatrix}
	\mathbb{s}_{i}^{Z}\\[2pt]\hline\\[-8pt] \mathbb{s}_{i}^{X}
	\end{pmatrix}	
\end{equation}
where $ \mathbb{s}_{i}^{Z} $ and $ \mathbb{s}_{i}^{X} $ are $ N\times 1 $ column vectors with elements $ \mathbb{s}_{i\mu}^{Z} $ and $ \mathbb{s}_{i\mu}^{X} $, respectively:
\begin{equation}
\begin{split}
	s_{i\mu}&=I:\;\left\lbrace  \begin{split}\mathbb{s}_{i\mu}^{Z}&=0\\\mathbb{s}_{i\mu}^{X}&=0\end{split}\right.\;\;,\;\;
	s_{i\mu}=X:\;\left\lbrace  \begin{split}\mathbb{s}_{i\mu}^{Z}&=0\\\mathbb{s}_{i\mu}^{X}&=1\end{split}\right. \\
	s_{i\mu}&=Y:\;\left\lbrace  \begin{split}\mathbb{s}_{i\mu}^{Z}&=1\\\mathbb{s}_{i\mu}^{X}&=1\end{split}\right.\;\;,\;\;
	s_{i\mu}=Z:\;\left\lbrace  \begin{split}\mathbb{s}_{i\mu}^{Z}&=1\\\mathbb{s}_{i\mu}^{X}&=0\end{split}\right. 
\end{split}
\end{equation}

The generator set $ S $ of the stabilizer state $ \ket{\mathcal{S}} $ is represented by the $ 2N\times N $ binary matrix $ \mathbb{S} $ where each column represents one stabilizer operator in $ S $:
\begin{equation}
\mathbb{S}=\left( \begin{split}&\mathbb{S}^{Z}\\[2pt]\hline\\[-12pt]&\mathbb{S}^{X}\end{split}\,\right)=\left( \begin{array}{ccccc}\mathbb{s}_{1}^{Z}&\cdots&\mathbb{s}_{i}^{Z}&\cdots&\mathbb{s}_{N}^{Z}\\[2pt]\hline\\[-8pt]\mathbb{s}_{1}^{X}&\cdots&\mathbb{s}_{i}^{X}&\cdots&\mathbb{s}_{N}^{X}\end{array}\,\right)	.
\end{equation}

For instance, for the natural generator set of a graph state like the one in Fig.~\ref{Graphs}(f) the upper block $ \mathbb{S}^{Z}=\varGamma $ is the adjacency matrix of the underlying graph, and the lower block $ \mathbb{S}^{X}=\mathbb{I}_{N} $ is the $ N\times N $ identity matrix. The upper and lower blocks for the seven-qubit color code generator set in Fig.~\ref{Stabilizer_state}(b) are respectively:
\begin{equation}
\mathbb{S}^{Z}=\begin{pmatrix}
1&0&0&0&0&0&0\\1&1&0&0&0&0&0\\1&1&1&0&0&0&0\\1&0&1&0&0&0&0\\0&1&0&0&0&0&0\\0&1&1&0&0&0&0\\0&0&1&0&0&0&0
\end{pmatrix} \;\;,\;\;  \mathbb{S}^{X}=\begin{pmatrix}
0&0&0&1&0&0&1\\0&0&0&1&1&0&1\\0&0&0&1&1&1&1\\0&0&0&1&0&1&1\\0&0&0&0&1&0&1\\0&0&0&0&1&1&1\\0&0&0&0&0&1&1
\end{pmatrix},	
\end{equation}
where we have written the $ Z $-type generators in the first three columns, and the $ X $-type stabilizers in the last four columns.

Two stabilizers $ s_{i},s_{j} $ commute if they satisfy that $ \mathbb{s}_{i}^{\scriptscriptstyle T}P\mathbb{s}_{j}=0 $ where $ P $ permutes the $ Z $ and $ X $ blocks:
\begin{equation}
P=\left( 
\begin{array}{c|c}
0\,&\mathbb{I}_{N}
\\ [2pt]\hline\\[-10pt]
\mathbb{I}_{N}\,&0
\end{array}\right)	.
\end{equation}

Then, the binary matrix that represents a generator set $ S $ must satisfy $ \mathbb{S}^{\scriptscriptstyle T}P\mathbb{S}=0 $, and consequently: $ \left[\mathbb{S}^{Z}\right] ^{\scriptscriptstyle T} \mathbb{S}^{X}+\left[ \mathbb{S}^{X}\right] ^{\scriptscriptstyle T}\mathbb{S}^{Z}=0 $.

A recombination of the stabilizers in the generator set $ S $ is represented in the binary picture as an invertible recombination of the columns of $ \mathbb{S} $ performed by a $ N\times N $ non-singular binary matrix $ R $. In the particular example of Eq.~(\ref{alt_basis}), the recombined binary matrix is $ \mathbb{S}^{(R)}=\mathbb{S}R $ where:
\begin{equation}
	R=\begin{pmatrix}
	1&0&0&0&0&0&0\\0&1&0&0&0&0&0\\0&0&1&0&0&0&0\\1&0&0&1&0&0&0\\0&1&0&0&1&0&0\\0&0&1&0&0&1&0\\0&0&0&0&0&0&1
	\end{pmatrix}	.
\end{equation}

Single-qubit local Clifford unitary operations are represented by the $ 2\times2 $ binary non-singular matrix
\begin{equation}
\mathbb{Q}=\left( 
\begin{array}{c|c}
a&b
\\ \hline
c&d
\end{array}\right)
\end{equation}
where the non-singularity in the binary picture means that the determinant calculated with the addition modulo 2 is non-zero: $ ad+bc=1 $ with $ a,b,c,d=0,1 $.

There are six of these matrices, each one corresponding to an invertible transformation of the Pauli operators. For instance, the Hadamard unitary $ H $ that performs the transformation: $ X\leftrightarrow Z $, $ Y\mapsto Y $, is represented by the binary matrix $ \mathbb{H} $:
\begin{equation}
\left\lbrace  \begin{split}
HXH^{\dagger}&=+Z\\
HYH^{\dagger}&=-Y\\
HZH^{\dagger}&=+X
\end{split}\right.\;,\; \left\lbrace  \begin{split}
\mathbb{H}&\begin{pmatrix}
0\\\hline 1
\end{pmatrix}=\begin{pmatrix}
1\\\hline 0
\end{pmatrix}\\
\mathbb{H}&\begin{pmatrix}
1\\\hline 1
\end{pmatrix}=\begin{pmatrix}
1\\\hline 1
\end{pmatrix}\\
\mathbb{H}&\begin{pmatrix}
1\\\hline 0
\end{pmatrix}=\begin{pmatrix}
0\\\hline 1
\end{pmatrix}
\end{split}\right.\;,\;\mathbb{H}=\left( 
\begin{array}{c|c}
0&1
\\ \hline
1&0
\end{array}\right)	.
\end{equation}

The binary picture disregards the phases; in fact $ \mathbb{H} $ represents any of the four Clifford operations $ Z^{\alpha}\mathrm{exp}\left( (-1)^{\beta}iY\pi/4\right)  $, where $ \alpha,\beta=0,1 $ ($ Z^{0}=I $) are two indices that control the four possible phases under this transformation of the Pauli operators. The six ways of transforming Pauli operators and the unitary representations of them are summarized in \cite{Hein2006}.

A local Clifford unitary operation on $ N $ qubits is represented by the $ 2N\times2N $ matrix
\begin{equation}
\mathbb{Q}_{L}=\left( 
\begin{array}{c|c}
A\,&B
\\\hline\\[-10pt]
C\,&D
\end{array}\right)	
\end{equation}
where $ A,B,C $ and $ D $ are the $ N\times N $ diagonal matrix that satisfy $ AD+BC=\mathbb{I}_{N} $ to guarantee that it preserves the commutation of Pauli operators: $ \mathbb{Q}_{L}^{\scriptscriptstyle T}P\mathbb{Q}_{L}=P $. Using this equality it is easy to see that the inverse local Clifford unitary is:
\begin{equation}
\mathbb{Q}_{L}^{-1}=\left( 
\begin{array}{c|c}
D\,&B
\\ \hline
C\,&A
\end{array}\right)	.
\end{equation}
In the binary picture, the problem of finding a graph state locally equivalent to the stabilizer state reduces to solving the following equation
\begin{equation}
	\mathbb{Q}_{L}\mathbb{S}R=\begin{pmatrix}
	\varGamma\\\hline \mathbb{I}_{N}
	\end{pmatrix}	
\end{equation}
where the variables are $ \mathbb{Q}_{L} $, $ R $ and $ \varGamma $.

Equivalently, the stabilizers represented by $ \mathbb{Q}_{L}\mathbb{S}R $ commute, so they satisfy that
\begin{equation}
\left( \varGamma\vert \mathbb{I}_{N}\right) P\mathbb{Q}_{L}\mathbb{S}R=0\;\;\;\Rightarrow\;\;\;\left( \varGamma\vert \mathbb{I}_{N}\right) P\mathbb{Q}_{L}\mathbb{S}=0	
\end{equation}
where now the variables are $ \mathbb{Q}_{L} $ and $ \varGamma $.

\section{Proof of Eq.~(\ref{pseudo_recombined})}\label{recombinations}
Here we prove Eq.~(\ref{pseudo_recombined}). The idea of the proof consists in writing $ M\left( W_{\varOmega}\right) $ in the binary representation and checking that an invertible recombination of $ W_{\varOmega}  $ results in an invertible recombination of the rows of $ M\left( W_{\varOmega}\right) $.

The representation $ \mathbb{W}_{\varOmega} $ of a generator subset $ W_{\varOmega} $ with $ n $ stabilizers is a $ 2N\times n $ binary matrix whose columns represent the stabilizers. For instance, the binary representation of the subset in Fig.~\ref{Direct_method}(a) is:
\begin{equation}
	\mathbb{W}_{\{2,3,4\}}=\begin{pmatrix}
	1&0&0\\1&0&0\\1&0&0\\1&0&0\\0&0&0\\0&0&0\\0&0&0\\\hline 
	0&0&0\\0&1&0\\0&1&1\\0&0&1\\0&1&0\\0&1&1\\0&0&1	
	\end{pmatrix}	.
\end{equation}

Given the binary representation $ \mathbb{s}_{i}\,,\mathbb{s}_{j} $ of two stabilizers $ s_i,\,s_j $ one can see that the pseudo-incidence matrix $ M\left( \{s_{i},s_{j}\}\right)  $ is the $ N\times1 $ column vector $ M\left( \{s_{i},s_{j}\}\right) $ with elements defined by:
\begin{equation}
	M\left( \{s_{i},s_{j}\}\right)_{\mu}=\mathbb{s}_{i\mu}^{Z}\mathbb{s}_{j\mu}^{X}+\mathbb{s}_{j\mu}^{Z}\mathbb{s}_{i\mu}^{X}	
\end{equation}
where the sum is performed modulo 2. Therefore, the elements of $ M\left( W_{\varOmega}\right)  $ are defined by:
\begin{equation}
\begin{split}
&M\left( W_{\varOmega}\right)_{\mu \{i,j\}}= \mathbb{W}_{\varOmega,i\mu}^{Z}\mathbb{W}_{\varOmega,j\mu}^{X}+\mathbb{W}_{\varOmega,j\mu}^{Z}\mathbb{W}_{\varOmega,i\mu}^{X}\\
&=\left( u_{\mu}^{\scriptscriptstyle T}\;|\;0\right) \mathbb{W}_{\varOmega}\left( v_{i}v_{j}^{\scriptscriptstyle T}+v_{j}v_{i}^{\scriptscriptstyle T}\right) \mathbb{W}_{\varOmega}^{\scriptscriptstyle T}\left( \begin{gathered}
0\\ \hline
u_{\mu}
\end{gathered}\right) 
\end{split}	
\end{equation}
where the column index $ l=\{i,j\} $ runs over all the pairs of stabilizers in $ W_{\varOmega} $. Here the auxiliary binary column vectors $ u_{\mu},v_{i} $ of size $ N $ and $ n $ respectively have one $ 1 $ entry in the position $ \mu=1,\ldots,N $ and $ i=1,\ldots,n $ respectively, and zeros elsewhere.

For simplicity, let us expand $ M\left( W_{\varOmega}\right) $ to have one column for every of the $ n^{2} $ possible pairs $ \{i,j\} $, including those with equal indices $ \{i,i\} $, and both orderings $ \{i,j\} $ and with $ \{j,i\} $ for the rest. Note that this expanded pseudo-incidence matrix has the same rank because the new columns $ \{i,i\} $ are defined as zero columns and the new columns $ \{j,i\} $ coincide with the existing columns $ \{i,j\} $. Thus, in the following, the indices $ i,j $ run over all $ n^{2} $ possibilities and hence, the pseudo-incidence matrix is a $ N\times n^{2} $ matrix. Moreover, a recombined $ W_{\varOmega}^{(R_{\varOmega})} $ is represented by $ \mathbb{W}_{\varOmega}R_{\varOmega} $, where $ R_{\varOmega} $ is a non-singular $ N\times n $ binary matrix, resulting in
\begin{equation}
\begin{split}
	&M\left(W_{\varOmega}^{(R_{\varOmega})}\right)_{\mu \{i,j\}}=\\
	&=\left( u_{\mu}^{\scriptscriptstyle T}\;|\;0\right) \mathbb{W}_{\varOmega}R_{\varOmega}\left( v_{i}v_{j}^{\scriptscriptstyle T}+v_{j}v_{i}^{\scriptscriptstyle T}\right)R_{\varOmega}^{\scriptscriptstyle T} \mathbb{W}_{\varOmega}^{\scriptscriptstyle T}\left( \begin{gathered}
0\\ \hline
u_{\mu}
\end{gathered}\right)
\end{split} 	.
\end{equation}
One can check that the terms in the center of this expression can be rewritten as
\begin{equation}
\begin{split}
	&R_{\varOmega}\left( v_{i}v_{j}^{\scriptscriptstyle T}+v_{j}v_{i}^{\scriptscriptstyle T}\right)R_{\varOmega}^{\scriptscriptstyle T}\\
	&=\sum_{a,b=1}^{n}\left( v_{a}v_{b}^{\scriptscriptstyle T}+v_{b}v_{a}^{\scriptscriptstyle T}\right)R_{\varOmega,ai}R_{\varOmega,bj}
\end{split}	
\end{equation}
where we have used that $ R_{\varOmega}=\sum_{a,b=1}^{n}R_{\varOmega,ab}v_{a}v_{b}^{\scriptscriptstyle T} $ and that $ v_{b}^{\scriptscriptstyle T}v_{i}=\delta_{bi} $.

The products $ R_{\varOmega,ai}R_{\varOmega,bj} $ are the matrix elements $ \tilde{R}_{\{a,b\}\{i,j\}} $ of a $ n^{2}\times n^{2} $ matrix $ \tilde{R}=R_{\varOmega}\times R_{\varOmega} $, which is non-singular because $ R_{\varOmega} $ is non-singular. This matrix performs a recombination of columns on the modified pseudo-incidence matrix, 
\begin{equation}\label{rank_preserved}
\begin{split}
&M\left( W_{\varOmega}^{(R_{\varOmega})}\right)_{\mu \{i,j\}}=\\
&=\sum_{\{a,b\}}\left( u_{\mu}^{\scriptscriptstyle T}\;|\;0\right) \mathbb{W}_{\varOmega}\left( v_{a}v_{b}^{\scriptscriptstyle T}+v_{a}v_{b}^{\scriptscriptstyle T}\right) \mathbb{W}_{\varOmega}^{\scriptscriptstyle T}\left( \begin{gathered}
0\\ \hline
u_{\mu}
\end{gathered}\right) \tilde{R}_{\{a,b\}\{i,j\}}\\
&=\sum_{\{a,b\}}M\left( W_{\varOmega}\right)_{\mu \{a,b\}}\tilde{R}_{\{a,b\}\{i,j\}}
\end{split}	.
\end{equation}
This completes the proof of Eq.~(\ref{pseudo_recombined}), namely that $ M\left( W_{\varOmega}(R_{\varOmega})\right)=M\left( W_{\varOmega}\right)\tilde{R}$.

\section{Example of a valid local witness that can not be found with the graph-based method}\label{non_graph-based_wit}
Here we present example of a local witness operator that can be found with the direct method, but not using the graph-based method. Both methods find generator subsets $ W_{\varOmega} $ that satisfy property $ (A) $ in Proposition \ref{p}. The graph-based method uses the local unitary operations $ U_{LE} $ that transform the stabilizer state into a graph state $ U_{LE}\ket{\mathcal{S}}=\ket{\mathcal{G}} $. These $ U_{LE} $ guarantee the existence of $ W_{\varOmega} $ because $ W_{\varOmega}^{\mathcal{G}} $ exists, 
\begin{equation}
	U_{LE}\ket{\mathcal{S}}=\ket{\mathcal{G}}\;\;\Rightarrow\;\;U_{LE}W_{\varOmega}U_{LE}^{\dagger}=W_{\varOmega}^{\mathcal{G}}	.
\end{equation}

On the other hand every generator subset $ W_{\varOmega} $ found with the direct method is directly built to satisfy property $ (A) $ in Proposition \ref{p}, i.e.~there exists a local unitary $ V_{LE} $ that transforms $ W_{\varOmega} $ into some $ W_{\varOmega}^{\mathcal{G}} $ (up to some recombination). But this does not imply that $ V_{LE} $ transforms the stabilizer state $ \ket{\mathcal{S}} $ into a graph state $ \ket{\mathcal{G}} $, 
 \begin{equation}
 V_{LE}\ket{\mathcal{S}}=\ket{\mathcal{G}}\;\;\nLeftarrow\;\;V_{LE}W_{\varOmega}V_{LE}^{\dagger}=W_{\varOmega}^{\mathcal{G}}	.
 \end{equation}
 
For instance, the subset $ W_{\{2,3,4\}} $ in Fig.~\ref{Direct_method}(a) is transformed into a generator subset $ W_{\varOmega}^{\mathcal{G}} $ with the local unitary $ V_{LE}=H_{3}H_{5}H_{6}H_{7} $, 
\begin{equation}
\begin{split}
	s_{R}^{Z}=Z_{1}Z_{2}Z_{3}Z_{4}&\mapsto g_{3}=Z_{1}Z_{2}X_{3}Z_{4},\\
	s_{B}^{X}=X_{2}X_{3}X_{5}X_{6}&\mapsto g_{2}=X_{2}Z_{3}Z_{5}Z_{6},\\
	s_{G}^{X}=X_{3}X_{4}X_{6}X_{7}&\mapsto g_{4}=Z_{3}X_{4}Z_{6}Z_{7},
\end{split}
\end{equation}
but $ V_{LE}\ket{\mathcal{S}} $ is not a graph state because the rest of generators in the basis $ S $ of Fig.~\ref{Stabilizer_state}(b) are not graph state generators under any recombination, 
\begin{equation}
	\begin{split}
	s_{R}^{X}=X_{1}X_{2}X_{3}X_{4}&\mapsto X_{1}X_{2}Z_{3}X_{4},\\
	s_{B}^{Z}=Z_{2}Z_{3}Z_{5}Z_{6}&\mapsto Z_{2}X_{3}X_{5}X_{6},\\
	s_{G}^{Z}=Z_{3}Z_{4}Z_{6}Z_{7}&\mapsto X_{3}Z_{4}X_{6}X_{7},\\
	s_{L}^{X}=X_{1}X_{2}X_{3}X_{4}X_{5}X_{6}X_{7}&\mapsto X_{1}X_{2}Z_{3}X_{4}Z_{5}Z_{6}Z_{7}.
	\end{split}
\end{equation}

Furthermore, we are going to prove here that for this particular generator subset $ W_{\{2,3,4\}} $ there is no $ V_{LE} $ that transforms $ \ket{\mathcal{S}} $ into some $ \ket{\mathcal{G}} $. In the binary picture it becomes clear that $ V_{LE} $ is partially fixed because it transforms $ W_{\{2,3,4\}} $ into some $ W_{\varOmega}^{\mathcal{G}} $. This fixing is enough to forbid the existence of any non-singular matrix $ R $ that recombines the color code generator set $ S $ into a recombined generator set $ S^{(R)} $ that transforms into the generator set $ G $ of some graph state under the action of $ V_{LE} $.
\\

First, let us show that the binary form of a generator subset $ W_{\varOmega} $ that satisfies property $ (A) $ in Proposition \ref{p}, or equivalently, that satisfies conditions $ (i),(ii),(iii),(iv) $ in Theorem \ref{t}, is the $ 2N\times n $ binary matrix $ \mathbb{W}_{\varOmega} $
\begin{equation}
\mathbb{W}_{\varOmega}=\left( \begin{gathered}
\mathbb{S}_{\varOmega}^{Z}\\
\varLambda^{Z}\varSigma_{\bar{\varOmega}}\\[2pt]\hline\\[-12pt]
\mathbb{S}_{\varOmega}^{X}\\
\varLambda^{X}\varSigma_{\bar{\varOmega}}
\end{gathered}\right) 	.
\end{equation}

The rows have been reordered to represent the qubits in $ \varOmega $ with the $ n\times n $ binary block matrices $ \mathbb{S}_{\varOmega}^{X}, \mathbb{S}_{\varOmega}^{Z} $, and the qubits in $ \bar{\varOmega} $ with the product of the $  (N-n)\times n $ matrix $ \varSigma_{\bar{\varOmega}} $ and the $ (N-n)\times (N-n) $ diagonal matrices $ \varLambda^{Z},\varLambda^{X}  $.

To satisfy $ (i) $, the rank modulo 2 of this matrix must be $ n $. To satisfy $ (ii) $ the rank modulo 2 of the following matrix, which represents the reduced Pauli operators $ s_{\varOmega,i} $, must be $ n $ as well, 
\begin{equation}\label{reduced_binary}
\left( \begin{gathered}
\mathbb{S}_{\varOmega}^{Z}\\[2pt]\hline\\[-12pt]
\mathbb{S}_{\varOmega}^{X}
\end{gathered}\right)	.
\end{equation}

The form of the blocks $ \varLambda^{Z}\varSigma_{\bar{\varOmega}} $ and $ \varLambda^{X}\varSigma_{\bar{\varOmega}} $ that represent the single-qubit Pauli operators in the qubits of $ \bar{\varOmega} $ guarantees that the stabilizers commute on each qubit of $ \bar{\varOmega} $, which is the condition $ (iii) $. To see this note that the matrices $ \varLambda^{Z},\varLambda^{X} $ are diagonal, so they just lead to certain rows of $ \varSigma_{\bar{\varOmega}} $ with all entries zero. For a qubit $ \mu\in\bar{\varOmega} $ the $ \mu $-th and $ (N+\mu) $th rows of $ \mathbb{W}_{\varOmega} $ can be of any of these four types, 
\begin{equation}
\left( \begin{gathered}\varSigma_{\bar{\varOmega},\nu}\\[2pt]\hline\\[-12pt]\varSigma_{\bar{\varOmega},\nu}\end{gathered}\right)\;\;,\;\;\left( \begin{gathered}\varSigma_{\bar{\varOmega},\nu}\\[2pt]\hline\\[-12pt]0\end{gathered}\right)\;\;,\;\;\left( \begin{gathered}0\\[2pt]\hline\\[-12pt]\varSigma_{\bar{\varOmega},\nu}\end{gathered}\right)\;\;,\;\;\left( \begin{gathered}0\\[2pt]\hline\\[-12pt] 0\end{gathered}\right),
\end{equation}
depending on the four combinations of $ 1 $ and $ 0 $ that $ \varLambda^{Z},\,\varLambda^{X} $ have in the diagonal position corresponding to the qubit $ \mu $. Here $ \varSigma_{\bar{\varOmega},\nu} $ is the $ \nu $-th row of $ \varSigma_{\bar{\varOmega}} $, with $ \nu=\mu-n $. For every stabilizer $ s_i\in W_\varOmega $, the first case indicates that the single-qubit Pauli operator $ s_{i\mu} $ is $ I $ or $ Y $, for the second case it is either $ I $ or $ Z $, for the third case $ I $ or $ X $ depending if $ \varSigma_{\bar{\varOmega},\mu i}=0 $ or $ 1 $ respectively, and for the fourth case $ s_{i\mu}=I $. Consequently, every stabilizer $ s_i\in W_\varOmega $ with support on qubit $ \mu\in\bar{\varOmega} $ commutes by construction, which is condition $ (iii) $.
\\

Now we focus on the particular generator subset $ W_{\{2,3,4\}} $. It contains $ n=3 $ independent stabilizers so it can be part of a generator set $ S $ represented by $ \mathbb{S} $ of the stabilizer state. We can then write its binary representation $ \mathbb{W}_{\{2,3,4\}} $ in the first three columns of $ \mathbb{S} $ and write the rest of generators in the other four columns, 
\begin{equation}\label{generator_set}
	\mathbb{S}=\left( \begin{array}{cc}
\mathbb{S}_{\varOmega}^{Z}&\tilde{\mathbb{S}}_{\varOmega}^{Z}\\[2pt]
\varLambda^{Z}\varSigma_{\bar{\varOmega}}&\tilde{\mathbb{S}}_{\bar{\varOmega}}^{Z}\\[2pt]\hline\\[-8pt]
\mathbb{S}_{\varOmega}^{X}&\tilde{\mathbb{S}}_{\varOmega}^{X}\\[2pt]
\varLambda^{X}\varSigma_{\bar{\varOmega}}&\tilde{\mathbb{S}}_{\bar{\varOmega}}^{X}
\end{array}\right) =\begin{pmatrix}
1&0&0&1&0&0&0\\1&0&0&1&1&0&0\\1&0&0&0&1&0&0\\1&0&0&0&0&0&0\\0&0&0&1&0&0&0\\0&0&0&1&1&0&0\\0&0&0&0&1&0&0 \\ \hline 0&1&0&0&0&1&1\\0&1&1&0&0&1&1\\0&0&1&0&0&1&1\\0&0&0&0&0&1&1\\0&1&0&0&0&0&1\\0&1&1&0&0&0&1\\0&0&1&0&0&0&1
	\end{pmatrix}.
\end{equation}
Note that we have reordered the rows to identify more easily the blocks. From the first to the last row, the order of the qubits that they represent is: $ 2,3,4,1,5,6,7 $, so the first three rows represent the qubits in $ \varOmega=\{2,3,4\} $ and the last four rows the rest of qubits. From the left column to the right column, the order of the stabilizers that they represent is $ s_{R}^{Z}\,,\,s_{B}^{X}\,,\,s_{G}^{X}\,,\,s_{B}^{Z}\,,\,s_{G}^{Z}\,,\,s_{R}^{X}\,,\,s_{L}^{X} $, so the first three columns represent the stabilizers in $ W_{\{2,3,4\}} $.

Here, the following blocks can be identified:
\begin{equation}
\begin{split}
	\varLambda^{Z}\varSigma_{\bar{\varOmega}}&=\begin{pmatrix}
	1&0&0\\0&0&0\\0&0&0\\0&0&0
	\end{pmatrix}\;\;\;\;\;\;\;\varLambda^{X}\varSigma_{\bar{\varOmega}}=\begin{pmatrix}
	0&0&0\\0&1&0\\0&1&1\\0&0&1
	\end{pmatrix}\\
	\tilde{\mathbb{S}}_{\bar{\varOmega}}^{Z}&=\begin{pmatrix}
0&0&0&0\\1&0&0&0\\1&1&0&0\\0&1&0&0
	\end{pmatrix}\;\;\;\;\;\;\;\;\,\tilde{\mathbb{S}}_{\bar{\varOmega}}^{X}=\begin{pmatrix}
0&0&1&1\\0&0&0&1\\0&0&0&1\\0&0&0&1
	\end{pmatrix}
\end{split}
\end{equation}

From the blocks $ \varLambda^{Z}\varSigma_{\bar{\varOmega}}, 	\varLambda^{X}\varSigma_{\bar{\varOmega}} $ we can read $ \varLambda^{Z},\varLambda^{X},\varSigma_{\bar{\varOmega}} $, 
\begin{equation}
	\begin{gathered}
	\varSigma_{\bar{\varOmega}}=\begin{pmatrix}
	1&0&0\\0&1&0\\0&1&1\\0&0&1
	\end{pmatrix},\\
	\varLambda^{Z}=\mathrm{diag}(1000),\;\;\;\;	\varLambda^{X}=\mathrm{diag}(0111).
	\end{gathered}
\end{equation}
The diagonal matrices fix partially the part $ V_{LE,\bar{\varOmega}} $ of the local Clifford unitary $ V_{LE}=V_{LE,\varOmega}V_{LE,\bar{\varOmega}} $ acting on the qubits of $ \bar{\varOmega} $. This part must transform all the single-qubit Pauli operators in $ \bar{\varOmega} $ into $ I $ or $ Z $ because the generator subsets $ W_{\varOmega}^{\mathcal{G}} $ have only $ I $ or $ Z $ on each qubit of $ \bar{\varOmega} $. To do so, the part $ V_{LE,\bar{\varOmega}} $ must be represented by
\begin{equation}
	\mathbb{Q}_{LE,\bar{\varOmega}}=\left( 
	\begin{array}{c|c}
	A\,&B
	\\[2pt]\hline\\[-8pt]
	\varLambda^{X}\,&\varLambda^{Z}
	\end{array}\right),
\end{equation}
where $ A,B $ are some $ (N-n)\times(N-n) $ binary diagonal matrices that satisfy $ A\varLambda^{Z}+B\varLambda^{X}=\mathbb{I}_{\bar{\varOmega}} $. It transforms all single-qubit Pauli operators in $ \bar{\varOmega} $ of the stabilizers in $ W_{\varOmega} $ into $ I $ or $ Z $, 
\begin{equation}
	\mathbb{Q}_{LE,\bar{\varOmega}}\left( \begin{gathered}
	\varLambda^{Z}\varSigma_{\bar{\varOmega}}\\[2pt]\hline\\[-12pt]
	\varLambda^{X}\varSigma_{\bar{\varOmega}}
	\end{gathered}\right)=\left( \begin{gathered}
	\varSigma_{\bar{\varOmega}}\\[2pt]\hline\\[-12pt]
	0
	\end{gathered}\right)	.
\end{equation}
The effect on the rest of stabilizers can be seen from their effect on the corresponding blocks, 
\begin{equation}
\mathbb{Q}_{LE,\bar{\varOmega}}\left( \begin{gathered}
\tilde{\mathbb{S}}_{\bar{\varOmega}}^{Z}\\[2pt]\hline\\[-12pt]
\tilde{\mathbb{S}}_{\bar{\varOmega}}^{X}
\end{gathered}\right)=\left( \begin{gathered}
A\tilde{\mathbb{S}}_{\bar{\varOmega}}^{Z}+B\tilde{\mathbb{S}}_{\bar{\varOmega}}^{X}\\[2pt]\hline\\[-12pt]
\varLambda^{X}\tilde{\mathbb{S}}_{\bar{\varOmega}}^{Z}+\varLambda^{Z}\tilde{\mathbb{S}}_{\bar{\varOmega}}^{X}
\end{gathered}\right)	.
\end{equation}
Therefore, the modified block is fixed by $ \varLambda^{X},\varLambda^{Z} $:
\begin{equation}
	\tilde{\mathbb{S}}_{\bar{\varOmega}}^{X\prime}\equiv\varLambda^{X}\tilde{\mathbb{S}}_{\bar{\varOmega}}^{Z}+\varLambda^{Z}\tilde{\mathbb{S}}_{\bar{\varOmega}}^{X}=\begin{pmatrix}
	0&0&1&1\\1&0&0&0\\1&1&0&0\\0&1&0&0
	\end{pmatrix},
\end{equation}
which, importantly, is not an invertible matrix.

We will now show that there is no local unitary $ V_{LE} $ represented by $ \mathbb{Q}_{LE} $ that transforms the generator set $ \mathbb{S} $ into graph state generators and also transforms $ \mathbb{W}_{\varOmega} $ into graph state generators. That means that $ \mathbb{Q}_{LE}\mathbb{S}R $ does not represent the natural generator set $ G $ of a graph state for any $ R $, 
\begin{equation}\label{graph_bin}
	\left( \begin{array}{cc}
	\varGamma_{\varOmega} & \gamma^{\scriptscriptstyle T}\\
	\gamma & \varGamma_{\bar{\varOmega}} \\[2pt]\hline\\[-12pt]
	\mathbb{I}_{n}&0\\
	0&\mathbb{I}_{(N-n)}
	\end{array}\right) 	.
\end{equation}
We can consider a completely general non-singular matrix $ R $ that performs the recombination in block form as
\begin{equation}
	R=\left( \begin{array}{cc}
	r_{\varOmega} & r_{\varOmega\bar{\varOmega}}\\
	r_{\bar{\varOmega}\varOmega} & r_{\bar{\varOmega}} 
	\end{array}\right).
\end{equation}
Here, $ r_{\varOmega},r_{\bar{\varOmega}} $ are $ n\times n $ and $ (N-n)\times(N-n) $ binary matrices respectively, and $ r_{\varOmega\bar{\varOmega}},r_{\bar{\varOmega}\varOmega} $ are $ n\times(N-n) $ and $ (N-n)\times n $ binary matrices respectively, such that $ R $ is non-singular.

Then, the lower blocks in Eq.~(\ref{generator_set}) change in this way:
\begin{equation}
\begin{split}
	&\mathbb{Q}_{LE,\bar{\varOmega}}\left(\varLambda^{X}\varSigma_{\bar{\varOmega}}\;\;\; \mathbb{S}_{\bar{\varOmega}}^{X\prime}\right)R=\left(0\;\;\; \tilde{\mathbb{S}}_{\bar{\varOmega}}^{X\prime}\right)R \\
	&=\left(\tilde{\mathbb{S}}_{\bar{\varOmega}}^{X\prime}r_{\bar{\varOmega}\varOmega}\;\;\; \tilde{\mathbb{S}}_{\bar{\varOmega}}^{X\prime}r_{\bar{\varOmega}}\right)	.	
\end{split}	
\end{equation}
The key is that the new block $ \tilde{\mathbb{S}}_{\bar{\varOmega}}^{X\prime}r_{\bar{\varOmega}} $ in the right should be $ \mathbb{I}_{(N-n)} $ to represent the generator set of a graph state. However, given that $ \tilde{\mathbb{S}}_{\bar{\varOmega}}^{X\prime} $ is not invertible, there is no matrix $ r_{\bar{\varOmega}\varOmega} $ such that
\begin{equation}
	\tilde{\mathbb{S}}_{\bar{\varOmega}}^{X\prime}r_{\bar{\varOmega}\varOmega}=\mathbb{I}_{\bar{\varOmega}}	.
\end{equation}
This proves that there is no local unitary $ V_{LE} $ that satisfies property $ (A) $ in Proposition \ref{p} for the generator subset $ W_{\{2,3,4\}} $ of Fig.~\ref{Direct_method} and transforms the stabilizer state $ \ket{\mathcal{S}} $ into a graph state. Therefore, this supposes an example of a local witness that can be found with the direct method but not with the graph-based method.

\section{Variance of the witnesses}\label{bin_distr}
Here we briefly discuss the variance of the witness expectation values, as computed from the experimentally measured stabilizer data. 

The estimator of the expectation value of a stabilizer operator is the mean value of the outcomes $ +1 $ and $ -1 $ when the operator is measured $ M $ times. We treat the experimental value of a stabilizer operator as a binomially distributed random variable with two possible values: $ +1 $ and $ -1 $ with probabilities $ p_{i} $ and $ 1-p_{i} $, so the expectation value is
\begin{equation}
\braket{s_{i}}=2p_{i}-1.
\end{equation}

The binomial distribution gives the probability of obtaining $ M_{i} $ times the result $ +1 $ in a sample of $ M $ repetitions, which is the one obtained with probability $ p_{i} $. The variance of binomial distributions is given by
\begin{equation}\label{Agc}
\sigma^{2}(M_{i})=Mp_{i}(1-p_{i})	. 
\end{equation}
The binomial variable $ \braket{s_{i}} $ is related to $ M_{i} $ by
\begin{equation}\label{Agd}
\begin{split}
\braket{s_{i}}=\frac{2M_{i}}{M}-1
\end{split}
\end{equation}
and therefore, the variance of a function $ f $ depending of $ M_{i} $ is related to $ \sigma^{2}(M_{i}) $ as
\begin{equation}\label{Age}
\sigma^{2}(f(M_{i}))=\left. \left[ \dfrac{\mathrm{d}f(M_{i})}{\mathrm{d}M_{i}}\right]^{2}\right|_{Mp_{i}} \sigma^{2}(M_{i})	.
\end{equation}
Simple algebra then yields the variance of the stabilizers
\begin{equation}\label{Agf}
\sigma^{2}\left(\braket{s_{i}}\right)=\frac{1}{M}\left( 1-\braket{s_{i}}^{2}\right)	.
\end{equation}
Since the variance of a sum of independent random variables is given by the sum of the variances of each variable, we can finally compute the variance of witness operators with the experimental value $ \braket{s_{i}} $ of the stabilizers involved in the witness and the number of times $ M $ that the measurement was done, 
\begin{align}
\sigma^{2}\left( \mathcal{W}\right)&=\frac{1}{M}\sum_{i=1}^{2^{N}-1}\frac{1- \braket{s_{i}}^{2}}{2^{2N}},
\\
\sigma^{2}\left(\mathcal{W}_{a}\right)&=\frac{1}{2M}\sum_{i=1}^{N}\left(1- \braket{s_{i}}^{2}\right),
\\
\begin{split}
\sigma^{2}\left( \mathcal{W}_{2m}\right)&= \frac{1}{M}\sum_{s_{i}^{X}\in \mathcal{S}}\frac{1- \braket{s_{i}^{X}}^{2}}{\left( \left| \mathcal{S}^{X} \right|+1\right) ^{2} }\\
&+\frac{1}{M}\sum_{s_{i}^{Z}\in \mathcal{S}}\frac{1- \braket{s_{i}^{Z}}^{2}}{\left( \left| \mathcal{S}^{Z} \right|+1\right) ^{2}},
\end{split}	
\end{align}
where $ \mathcal{S}^{X} $ are the $ X $-type stabilizers $ s_{i}^{X}\in\mathcal{S} $ in the stabilizer $ \mathcal{S} $, and $ \mathcal{S}^{Z} $ are the $ Z $-type stabilizers $ s_{i}^{Z}\in \mathcal{S} $.
Finally, the variances of local witnesses constructed from the generator subset $ W_{\varOmega} $ are then given by
\begin{align}
\sigma^{2}\left( \mathcal{W}_{\varOmega}\right)&=\frac{1}{M}\sum_{s_{i}\in \left[ W_{\varOmega}\right] }\frac{1- \braket{s_{i}}^{2}}{2^{2n}},
\\
\sigma^{2}\left(\mathcal{W}_{\varOmega,a}\right)&=\frac{1}{2M}\sum_{s_{i}\in  W_{\varOmega}}\left(1- \braket{s_{i}}^{2}\right),
\\
\begin{split}
\sigma^{2}\left( \mathcal{W}_{\varOmega,2m}\right)&= \frac{1}{M}\sum_{s_{i}^{X}\in \left[  W_{\varOmega}\right] }\frac{1- \braket{s_{i}^{X}}^{2}}{\left( \left| \left[  W_{\varOmega}\right]  ^{X}\right|+1\right) ^{2}}\\
&+\frac{1}{M}\sum_{s_{i}^{Z}\in \left[  W_{\varOmega}\right] }\frac{1- \braket{s_{i}^{Z}}^{2}}{\left( \left| \left[   W_{\varOmega}\right]  ^{Z}\right|+1\right)^{2}},
\end{split}	
\end{align}
where $ \left[  W_{\varOmega}\right]^{X} ,\left[  W_{\varOmega}\right]^{Z}  $ are the $ X $-type and $ Z $-type stabilizers $ s_{i}^{X},s_{i}^{Z} $ in the spanned group $ \left[  W_{\varOmega}\right]  $ respectively, and the cardinality of the sets is $ \left| \left[  W_{\varOmega}\right]  ^{X}\right|,\left| \left[  W_{\varOmega}\right]  ^{Z}\right| $, respectively.

\bibliography{library}{}
\bibliographystyle{apsrev4-1}

\end{document}